\documentclass[english,graybox]{svmult}
\usepackage{mathptmx}
\usepackage{helvet}
\usepackage{courier}
\usepackage[T1]{fontenc}
\usepackage[latin9]{inputenc}
\usepackage{verbatim}
\usepackage{float}
\usepackage{url}
\usepackage{amsmath}
\usepackage{graphicx}
\usepackage{txfonts}
\usepackage[round]{natbib}

\makeatletter

\renewcommand{\cite}{\citet}

\usepackage{type1cm}

\usepackage{multicol} 
\usepackage[bottom]{footmisc} 

\makeatother

\usepackage{babel}

\def\kms{km ${\rm s}^{-1}$}
\def\sn1{${\rm s}^{-1}$}

\def\cmn{${\rm cm}^{-2}$}
\def\ergs{erg ${\rm s}^{-1}$}

\def\xmm{{\it XMM-Newton}}
\def\chan{{\it Chandra}}

\def\sp{\space}

\def\sgr{V4743 Sgr}
\def\xmm{{\it XMM-Newton}}
\def\swift{{\it SWIFT}}
\def\sun{{\odot}}
\def\apj{Astrophysical Journal}%
\def\apjl{Astrophysical Journal Letters}%

\newcommand\aj{{Astronomical Journal}}%
\newcommand\apjs{{Astrophysical Journal, Supplement}}%
\newcommand\apss{{Ap\&SS}}%
\newcommand\aap{{A\&A}}%
\newcommand\mnras{{MNRAS}}%
\newcommand\pasj{{PASJ}}%
\newcommand\nat{{Nature}}%
\begin{document}

\title*{Astrophysical Fluids of Novae: High Resolution Pre-decay X-ray spectrum of V4743 Sagittarii}

\titlerunning{Astrophysical Fluids of Novae}

\author{J.M Ram{\'i}rez-Velasquez}

\authorrunning{Ram{\'i}rez-Velasquez}

\institute{J.M Ram{\'i}rez-Velasquez \at Physics Centre, Venezuelan Institute for Scientific Research ({\sc ivic}), PO Box 20632, Caracas 1020A, Venezuela - 
\email{josem@ivic.gob.ve}
\and
Departamento de Matem\'aticas, {\sc cinvestav} del I.P.N., 
07360 M\'exico, D.F., M\'exico.}

\maketitle

\abstract{
Eight X-ray observations of V4743\,Sgr (2002), observed
with \chan \space and \xmm, are presented, covering
three phases: Early optically thin hard emission (day 50.2), photospheric
emission from the ejecta
(days 180.4, 196.1, 301.9, 371, 526), and faint post-outburst
emission (days 742 and 1286). The flux level at Earth during the first
and last phase is of order $10^{-12}$\,erg\,cm$^{-2}$\,s$^{-1}$
over the energy range 0.3-2.5\,keV. These values are higher than an
upper limit obtained in September 1990 with ROSAT. The nova thus
continued fading in the soft band (0.1-2.4\,keV).
The nova turned off some time between days 301.9 and 371, and
the X-ray flux subsequently decreased from day 301.9 to 526,
following an exponential decline time scale of $(96 \pm 3)$ days.
We use the absorption lines present in the SSS spectrum
for diagnostic purposes, and characterize the physics
and the dynamics of the expanding atmosphere during the
explosion of the nova. The information extracted from this first
stage is then used as input for computing full photoionization
models of the ejecta in V4743\,Sgr.
The SSS spectrum is modeled with a simple black-body
and multiplicative Gaussian lines, which provides us
of a general kinematical picture of the system,
before it decays to its
faint phase (Ness et al. 2003).
In the grating spectra taken between days 180.4 and 370, we can resolve
the line profiles of absorption lines arising from H-like and He-like
C, N, and O, including transitions involving higher principal
quantum numbers. Except for a few interstellar lines, all lines are
significantly blue-shifted, yielding velocities between 1000 and
6000 \kms \sp which implies an ongoing mass loss.
It is shown that significant expansion and mass loss
occur during this phase of the explosion, at a rate
$\dot{M} \approx (3-5) \times 10^{-4} ~ (\frac{L}{L_{38}}) ~ M_{\odot}/yr$.
Our measurements show that the efficiency of the amount of energy used 
for the motion of the ejecta,
defined as the ratio between the kinetic luminosity $L_{\rm kin}$ and 
the radiated luminosity
$L_{\rm rad}$, is of the order of one.
}

\section{Introduction}

Classical Novae (CNe) are the historically longest-known eruptive
sources in the sky. They occur in binary systems consisting of a
white dwarf (WD) and a normal star with typical orbital periods of
a few hours. Mass lost by the star leads to accumulation of a
hydrogen-rich layer on the WD surface that erupts in a
thermonuclear explosion after nuclear
ignition conditions are reached (\cite{st08}). The radiative
energy output of the outburst is large enough to eject both
accreted material and some dredged up WD material into space.
This envelope resembles a stellar atmosphere of an F giant
(\cite{novae64}). 
Early in the evolution, the ejecta are
bright in the optical, and no high-energy radiation is expected.
As the expansion continues, the
density drops, the photosphere moves inward in mass, exposing
successively
hotter plasma, and the effective temperature rises. This
process leads to the commonly observed decline in optical, while
the peak of the spectral energy distribution (SED) moves to higher
energies.
In this standard picture, at some point in the evolution the peak is
expected to reach the X-ray regime, and several novae have been observed
to emit a supersoft X-ray spectrum, similar to those of the
class of supersoft X-ray binary sources (SSS) such as Cal\,83
(\cite{starr04}).
These are extremely luminous sources ($\log L_{\rm bol}\approx
37$) with effective temperatures $\sim 30-40$\,eV ($T_{\rm eff}\approx
(3-5)\times 10^5$\,K). Such high
luminosities and temperatures, if observed in systems with WDs,
can only be powered by nuclear burning, and when
a nova is emitting an SSS spectrum, nuclear burning near the surface
of the WD must still be
continuing. The time at which a nova ceases to emit an SSS
spectrum can be considered as the time when nuclear burning has turned
off.

The first detailed X-ray study of a nova during outburst was
carried out for V1974\,Cyg (\cite{krautter1996a}).
While X-ray emission was not expected to be observable before the
photosphere receded to sufficiently hot layers within the envelope,
V1974 Cyg exhibited an early hard X-ray spectrum, apparently with the
characteristics of a collisional, optically-thin plasma 
(\cite{krautter1996a,balman1998a}).
Such early X-ray emission has now been seen in other novae, e.g.,
V838\,Her (\cite{lloyd92}), V382\,Vel (\cite{Orio2001}), several
CNe in a sample of novae observed with \swift \sp
(\cite{swnovae}), and,
recently, V2491\,Cyg
(\cite{page09}), and V458~Vul
(\cite{v458}).

X-ray observations during the SSS phase yield a physical
description of the innermost regions. X-ray spectra yielding
color-equivalent temperatures in excess of $\sim (2-3)\times10^5$\,K
imply that nuclear burning is the
energy source (\cite{ness_v723}).
The evolution of the SSS is governed by the evolution of the
nuclear-burning envelope that remains on top of the WD. Models of
post-outburst WD envelopes with steady H-burning show that
envelopes with smaller masses have smaller photopsheric
radii and higher effective temperatures.
As the H-rich envelope mass decreases due to mass loss
and conversion of H to He, the effective temperature is predicted
to increase at constant bolometric luminosity, reaching maximum
temperature just before turn-off.
This has recently been observed by \cite{page09} in V2491~Cyg.
The evolution of the effective temperature is faster at low
temperatures, slowing down at high temperatures, with its pace
depending on the WD mass and the particular envelope
composition (\cite{sala05a}).
For data with sufficient signal-to-noise ratio and spectral
resolution, non-LTE atmosphere models allow
quantitative conclusions. However, \cite{ness09} found
significant blue shifts in all available high-resolution X-ray
spectra of novae, indicating continuing expansion
and probably mass loss, such that rigorous application of
such methods must include the expansion
(\cite{vanrossumness09}).

An alternative route is the use of expanding photoionized envelope
models,
for the characterization of the main kinematics and
physical parameters of the system. And that is the main
goal of the work we introduce in this chapter.
Here, we present an analysis of the X-ray spectra from
the \chan \space
and \xmm \space
monitoring observations of
V4743~Sgr (2002) that were carried out between days 50.2 and 1286
after discovery. 


\section{Observations}
\label{obssect}

Altogether, five observations with \chan \space
and three with \xmm \space were obtained between days 50
and 1286 after outburst, probing different phases of the evolution.
In Table~\ref{obs} we list the journal of observations with
day after discovery.

\begin{table*}
\begin{flushleft}
\renewcommand{\arraystretch}{1.1}
\caption{\label{obs}Observation summary}
{
\begin{tabular}{llrllrrr}
\hline
\multicolumn{2}{l}{day$^a$\hfill ObsID} & Detector/mode & \multicolumn{2}{l}{Start\hfill Exp.\,time} &count&flux$^h$&flux$^i$\\
&&&date&\multicolumn{2}{l}{(ks)\hfill rate}&&\\
\hline
50.23 & 3774 & ACIS-S$^b$/FAINT & 2002-11-09.24 & 5.30 & 0.3$^g$ & $1.8\,\pm\,0.6$ & $1.3\,\pm\,0.2$\\
180.39$^d$ & 3775 & HRC$^b$/LETGS & 2003-03-19.40 & 14.7 & 44.0$^e$ & $2022\,\pm\,450$ & $1815\,\pm\,400$\\
180.56$^d$ & 3775 & HRC$^b$/LETGS & 2003-03-19.55 & 5.6 & 0.62$^e$ & $32\,\pm\,15$ & $24.1\,\pm\,8$\\
196.14 & \multicolumn{2}{l}{0127720501 \hfill RGS$^c$} & 2003-04-04.93 & 35.2 & 51.9$^f$ & $1709\,\pm\,573$ & $1709\,\pm\,573$\\
301.88 & 3776 & HRC$^b$/LETGS & 2003-07-18.90 & 11.7 & 43.5$^e$ & $2006\,\pm\,190$ & $1810\,\pm\,170$\\
370.98 & 4435 & HRC$^b$/LETGS & 2003-09-25.99 & 12.0 & 22.6$^e$ & $1044\,\pm\,350$ & $928\,\pm\,300$\\
526.05 & 5292 & HRC$^b$/LETGS & 2004-02-28.06 & 10.3 & 5.05$^e$ & $230\,\pm\,40$ & $172\,\pm\,30$\\
741.98 & \multicolumn{2}{l}{0204690101 \hfill MOS1$^c$/THIN} & 2004-09-30.77 & 22.1&0.086$^g$ & $2.6\,\pm\,1.0$ & $0.6\,\pm\,0.3$\\
1285.9 & \multicolumn{2}{l}{0304720101 \hfill MOS1$^c$/THIN} & 2006-03-28.64 & 34.1&0.055$^g$ & $1.5\,\pm\,0.5$ & $0.4\,\pm\,0.2$\\
\hline
\end{tabular}
}

$^a$after discovery (2002, September 19.8)\\
$^b$\chan \\
$^c$\xmm \sp (combined RGS1 and RGS2)\\
$^d$observation split in high-state and low-state\\
$^e$counts per second from dispersed photon in range $5-55$\,\AA\\
$^f$counts per second from dispersed photon in range $5-38$\,\AA\ (RGS1)\\
$^g$counts per second ($0.2-10$\,keV $=1.2-62$\,\AA)\\
$^h$absorbed band flux [ranges as in footnotes $e-g$] in units $10^{-12}\,$erg\,cm$^{-2}$s$^{-1}$; error ranges represent peak to peak variations\\
$^i$absorbed band flux [range $5-38$\,\AA $=0.3-2.5$\,keV] in units $10^{-12}\,$erg\,cm$^{-2}$s$^{-1}$; error ranges represent peak to peak variations

\renewcommand{\arraystretch}{1}
\end{flushleft}
\end{table*}


\subsection{Observing instruments}
\label{tech:instruments}

The ACIS-S is an array of
CCD chips, sensitive between $0.1-10$\,keV with an effective
area ranging from $110-600$\,cm$^2$ from $0.5-1.5$\,keV and an
energy resolution of roughly 100\,eV. The HRC
(High-Resolution-Camera) is a microchannel plate imaging detector
that records the dispersed photons of the Low Energy Transmission
Grating Spectrometer (LETGS). The LETGS covers a wavelength
range of $1-170$\,\AA\ with an effective area of $10-30$\,cm$^2$
in the $19-40$\,\AA\ range and a wavelength resolution of 0.05\,\AA.
Since the dispersion angle is proportional to wavelength, the
grating spectra are extracted in wavelength units. The Reflection
Grating Spectrometer (RGS) on board
\xmm \space
is
sensitive between $1-38$\,\AA\ with an effective area between
$20-60$\,cm$^2$ in the $19-40$\,\AA\ range and a wavelength
resolution of 0.05\,\AA.
The European Photon Imaging Camera (EPIC) consists of two
MOS (Metal Oxide Semi-conductor) CCD
detectors and a pn detector.
The MOS are sensitive
between 0.2-10\,keV with an effective area ranging between
$100-600$\,cm$^2$ from $0.5-1.5$\,keV and an energy resolution
of roughly 50\,eV.

\subsection{Extraction procedures}
\label{tech:extraction}

The main observation product follows the concept of events
files, which are tables of events with columns containing
positions, arrival times and energies for each individual
event. We generated these files from the raw products using
standard reduction routines provided by the respective
\chan \space CIAO \footnote{\tt http://cxc.harvard.edu/ciao} (v4.2) software and 
\xmm \space SAS (9.0.0).
We extracted average count rates and spectra from the events files
by application of filter criteria on photon positions, arrival times,
and energy (see below). In the right part of Table~\ref{obs} we give
the count rates and absorbed X-ray band fluxes for the energy ranges
indicated in the corresponding footnotes. Absorbed X-ray band fluxes
can be obtained from spectral models with fairly good accuracy. As
long as the spectra are reasonably well reproduced, the resultant
flux does not depend strongly on the model assumptions, and the
determination of absorbed fluxes over the energy band covered by the
detector is thus robust. Nevertheless, for the grating
spectra we made use of the high spectral resolution and determined
the fluxes directly from the spectra by integrating the photon
fluxes from each spectral bin in a given wavelength range.
First, we extracted the fluxes from the entire band pass of each
instrument, except for the LETGS fluxes, where significant
contamination of second-order photons occurs longwards of 55\,\AA.
For direct comparison of the brightness evolution, we also extracted
fluxes over a common band pass for all observations. Since the light
curves are highly variable, we consider the range between minimum
and maximum count rates as uncertainty ranges. The extracted fluxes
are listed in the last two columns of Table~\ref{obs}.

The first X-ray observation was made with low (CCD) spectral
resolution with the original purpose to determine whether the
brightness level was suitable for high-resolution observations.
We have placed a circular extraction region with a radius of
20 pixels around the expected source position and extracted the
number of counts in the detect cell. The background was
extracted from an adjacent (source-free) region. As can be seen
from Table~\ref{obs}, the source was clearly detected. Next, we
extracted a spectrum from the photon energies.
The spectral analysis is described in \S\ref{m:early},
and the fluxes listed in the last two columns of Table~\ref{obs}
have been directly integrated from the best-fit model.

The time period between days 50.2 and 180.4 was
unfortunately not covered with any observations because
Sagittarius was behind the Sun. As soon as V4743\,Sgr was
visible again, a bright SSS spectrum was observed by 
\chan \space
with
high-amplitude oscillations and a sudden, unexpected decay
by two orders of magnitude (\cite{ness2003a}). We split this observation
into two parts (starting on days 180.4 and 180.6) in order to
investigate the bright and faint phases separately.
Three
more
\chan \space
LETGS observations were taken on days 301.9, 371, and 526.
The spectrum was obtained by
following the standard
{\it Science Threads}
\footnote{\tt http://cxc.harvard.edu/ciao/threads/spectra\_letghrcs/}
for extraction of
LETG/HRC-S Grating spectra.
\footnote{Along with the calibration data CALDB V 4.2}
We use the new spectral extraction region
(i.e., a ``bow-tie" shaped region), contained
in CALDB 4.2.
When the LETG
({\it Low Energy Transmission
Grating}) is used with the HRC-S detector
({\it High Resolution Camera}), this
comprises a central rectangle
abutted to outer regions whose widths increase
as the dispersion distance increases.
The background region is taken from
above and below the dispersed spectra.
The region shape for both, source and background
negative and positive orders is precisely given
in the CALDB V 4.2.
Having properly extracted both, source and background
spectra from each arm, we merged them, obtaining added
source and background spectra. This is intended
to increase the
signal-to-noise ratio
(S/N) of the final spectrum, and the spectral
analysis (throughout this work) is based on this
co-added spectrum.

The decay phase, which comprises
the time interval before the faint-phase (obs day 180.6), of
approximately 5 ks of duration, is discussed in
Ness et al. (in prep.), and will not be analysed here.
The pre-decay phase
is composed then for the first 15 ks
of the observation at day 180.4, which we call epoch 1.
Due to similarities with this we also study the spectra
at days 301 and 370 called epochs 2 and 3, respectively.
The LETGS has an energy coverage range $0.07-10$ keV,
however,
we make use of the energy bandpass $0.2-10$ keV,
since significant contamination of second and higher 
orders photons are expected below 0.2 keV.
The final spectra is shown using 
a
bin size
of 0.0250 \AA \space for epochs 1, 2, and
the default bin size
of the LEG
0.0125 \AA \space for epoch 3, making use of the high resolution
of the spectrometer.


On day 196.1, two weeks after the observation of the steep decay,
an \xmm \space
ToO observation was carried out,
and the nova was again bright in X-rays exhibiting a SSS spectrum.
The spectra and count rates
were extracted from standard pipeline products. We have corrected the
RGS spectra for pile up by reclaiming the first-order photons that
were recorded in the second- and third-order RGS spectra.
For the calculation of
fluxes, we use the ``fluxed'' spectrum that
combines RGS1 and RGS2 spectra. The ``fluxed'' RGS spectrum is
a product of a SAS procedure that fully exploits the redundancy
of the two spectrometers onboard \xmm.

Two more observations were taken, on days 742 and 1286, with
\xmm. The RGS spectra do not contain sufficient signal for useful
results, and we concentrate on the MOS1 observations.
The source is not recorded in the pn detector because the source
position coincided with a gap between two CCD chips. The
count rates and spectra were extracted from the events file
using circular extraction regions with a radius of 200 pixels,
again filtering on the coordinates.

\subsection{Brightness evolution}
\label{bevol}

\begin{figure}
\begin{center}
\resizebox{6cm}{!}{\includegraphics[angle=0]{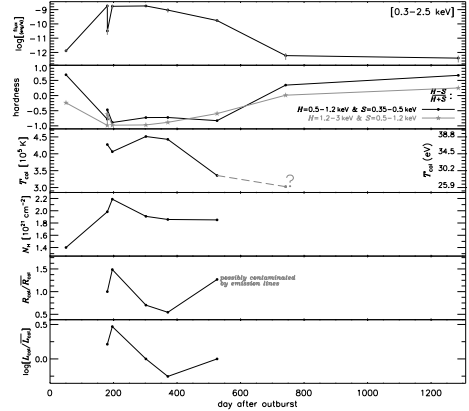}}
\caption{
Evolution of X-ray band fluxes from the last
column of Table~\ref{obs} (top) and hardness ratios (second panel).
The next four panels show the evolution of color-equivalent
temperature,
neutral hydrogen column density, $N_{\rm H}$, effective
radius, and bolometric luminosity. Only relative changes of radii and
luminosities are shown because the absolute values are unreliable
as a consequence of poor model assumptions. The value of the blackbody
temperature found from the model to day 742.
The Galactic absorption $N_{\rm H}$ obtained from the model to day 50.2
as described in
\S\ref{m:early} are included in the third and fourth panels,
respectively. Since the blackbody temperature for day 742 is
highly uncertain, it is marked with a gray question mark.
The normalization for day 526 might be contaminated by emission lines,
leading to an overestimated radius (5th panel).
\label{evol1}}
\end{center}
\end{figure}

The evolution of the X-ray broad-band fluxes is depicted in the top
panel of Fig.~\ref{evol1}. From day 50.2 to 180.6, the flux
increased by three orders of magnitude. The faint phase on day
180.6 yields a factor $\sim 75$ reduced brightness compared to
before the decay, but the flux level is still a factor $\sim 20$
higher than the flux obtained from the pre-SSS observation taken
on day 50.2 (see also Table~\ref{obs}). The flux on day 196.1 is
slightly lower than the pre-decay flux, while we find the same flux
level for day 301.9 as before the decay. We caution that the comparison
of fluxes between days 180 and 196.1 is uncertain owing to
cross-calibration uncertainties between different instruments and
uncertainties arising from the pile up correction of the RGS spectrum.
Note that the \chan \space LETG spectra are not piled up because of
the architecture of the HRC detector.
The fluxes for days 301.9, 371, and 526.1
follow an exponential decline, e$^{-t/\alpha}$, with
$\alpha=(96\,\pm\,3)$\,days (1-$\sigma$ uncertainty).

The last two \xmm \space observations taken on days 741 and 1286 yield
fairly similar flux levels as on day 50.2. In order to determine whether
this may be the quiescent level, we checked the ROSAT archive
and found no X-ray detection in a 12-ks observation
(ObsID 932149) taken 1990 September 17.1. The upper limit of
$2\times10^{-4}$\,cps over the 0.1-2.4 bandpass corresponds to a
\chan \space and \xmm \space 
count rate of $<10^{-3}$ cps and is thus well
below the count rate level observed on days 50.2, 741, and 1286
encountered for the same band.

\subsection{Spectral Evolution}
\label{sevol}

As a first description of the spectral characteristics of the object, we extracted
count rates from different energy ranges, and calculated two
hardness ratios along the conventional definition $HR=(H-S)/(H+S)$
with $H$ and $S$ being the count rates in two hard and soft bands,
respectively. The evolution of the two different hardness ratios
are shown in the second panel of Fig.~\ref{evol1}, where the
energy ranges for $H$ and $S$ are indicated in the bottom right
legend. The evolution of both hardness ratios demonstrates the
three phases of evolution. We group the observations into an early hard spectrum
(day 50), the Super Soft Source phase
(days 180-370), and the post-SSS phase (days 526-1285).
The panels below show the evolution of the spectral shape
of the continuum color and
related parameters characterizing the SSS spectra.

\begin{figure}
\begin{center}
\resizebox{6cm}{!}{\includegraphics[angle=0]{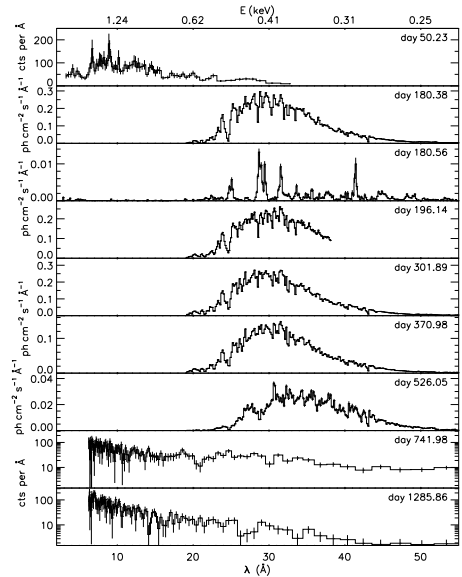}}
\caption{
Comparison of all X-ray spectra with the
start times given as fractional days after discovery in the upper
right legend of each panel. For observation
details see Table~\ref{obs}. The grating spectra were converted to
photon fluxes (days 180.4-526), and data shortwards of 19\,\AA \sp
were removed as no signal is present.
The CCD spectra (days 50.2, 724, and 1286) have
been extracted on an equidistant energy grid, but are shown on the
same wavelength scale, and in units counts per \AA, without background
subtraction. In the bottom two panels, the spectra are shown on a
logarithmic scale.
\label{cmpspec}}
\end{center}
\end{figure}

The detailed spectra are shown in Fig.~\ref{cmpspec} in the
order of date after discovery from top to bottom (see upper right
legends in each panel). While the grating spectra taken between
days 180.4 and 526 can be converted to photon flux spectra,
this is not possible for the CCD spectra, and the raw counts
per bin are plotted instead. Since the CCD spectra are binned
on an equidistant {\em energy} grid, these spectra are not equidistantly
binned when plotted in wavelength units, which is chosen here
for consistency with the majority of grating spectra. For
orientation purposes, the corresponding energies to each given
wavelength label on the x-axis are given in the top.

The spectral evolution of novae in X-rays
can be divided into three major phases, which can be identified
in Fig.~\ref{cmpspec}: an early emission phase that is characterized
by ''hard`` emission (day 50.2), the SSS phase (days 180.4 to 526), and
a late emission phase that is again a harder spectrum (days 742 and
1286). While the SSS emission is photospheric emission from the
extended WD, the origin of X-ray emission during the other two phases is
less clear. Early ''hard`` emission is suspected to originate in
shocks while the late emission phase could be radiatively cooling
nebular ejecta.

The spectrum on day 50.2 is shown in the top panel of
Fig.~\ref{cmpspec}. It has very low signal to noise and spectral
resolution but significant hard and soft emission can clearly be
identified. Some of the features could be emission lines,
and this spectrum thus meets expectations of an early hard emission
line spectrum.

The LETGS spectrum taken on day 180.4 has already been presented
by \cite{ness2003a}. The spectrum is clearly significantly softer,
and it is a continuum spectrum with absorption lines. On day
180.6, after the steep decay, an emission line spectrum has 
appeared (\cite{ness2003a}). As described above, units of photon fluxes
are plotted.

The spectra taken on days 196, 302, 371, and 526 are remarkably
similar, and they all originate from the extended WD photosphere.
Small differences can be seen in the depth of some absorption lines,
and the Wien tail extends to somewhat shorter wavelengths between
days 196 and 372, compared to days 180.6 and 526. On day 526,
the absorption lines are less deep, and emission line features
arise. These emission lines could have been present all the time,
and with the lower emission level of the continuum on day 526, they
are easier to see. A detailed comparison of the SSS spectra can
be found in \S\ref{discu1}.

The last two \xmm \space
observations are too faint for the high-resolution
RGS, and only the low-resolution MOS1 spectra are useful.
In the bottom two panels of Fig.~\ref{cmpspec}, the spectra
from days 742 and 1286 are shown. Both spectra show no features
that could be associated with emission lines like in the ACIS
spectrum in the top panel of Fig.~\ref{cmpspec}. Since the MOS1
has a similar resolution as the ACIS, we can conclude that the
post-outburst spectra are of a different nature.


\subsubsection{Continuum and detection of X-ray lines in the SSS spectrum
\label{cont1}} 

Here we describe the selection of the continuum and the
statistical procedure for the detection of the X-ray lines. 
Motivated by Ness etal. 2010, we choose a black-body with temperature
$kT=37$ eV ($T=427000$ K).
The column density of neutral hydrogen in the line of sight,
$N_{\rm H}$, was found by comparison
of the non-absorbed model with $N_{\rm H}$-corrected observed
spectra, 
from different assumed values of $N_{\rm H}$.
Other authors argued that $\chi2$ minimization of
$N_{\rm H}$ overemphasizes the $25-35$-\AA\ region where the highest
count rates are encountered, while the importance of $N_{\rm H}$
increases towards longer wavelengths, where the count rate is low,
and the contributions to $\chi2$ are small.
We use the {\tt tbabs}
\footnote{The T\"ubingen-Boulder ISM absorption model}
model for the interstellar absorption with
a column density of
$N_H=2.2\times 10^{21}$ \cmn.
For the continuum described above we look for significant
residuals using as general strategy model comparison.
The resulting value
of $2.2\times10^{21}$\,cm$^{-2}$
is between
expected values from Galactic neutral hydrogen maps.
For the purposes of detecting a spectral feature,
i.e., an absorption/emission line 
(in the count-wavelength space)
is at least
two {\it changepoints} (separated by at least a
distance equivalent to the resolution
of the instrument). In the statistics literature this is
a point where the underlying process changes abruptly. The concept,
has already been used in time series data
(\cite{scargle2004a}).
Here we apply the idea to grating X-ray data.
So we take a ``chunk" of data, labeled $k$, and compute
the marginalized likelihood of the model ($M_k$)
given the data ($D_k$):

\begin{equation}
P(M_k | D_k)=P_0 \frac{\Gamma(N_k+\alpha)}
{(\sum \Lambda_n + \beta) ^{N_k+\alpha}},
\end{equation}
where $N_k$ is the total counts in the chunk of data $k$.
$\Lambda_n$ is the
broad band model count rate in units of counts per bin ($n$).
$P_0$ forms part of the prior:

\begin{equation}
P_0 l^{\alpha-1}\exp{(-\beta l)},
\end{equation}
where
\begin{equation}
P_0=\frac{\beta^{\alpha}}{\Gamma(\alpha)},
\end{equation}
and
$\Gamma$ is the incomplete gamma function, and $\beta$ and $\alpha$ are
parameters.

Our fitting of the continuum model into the data is made using
the Interactive Spectral Interpretation System
\footnote{{\tt http://space.mit.edu/CXC/ISIS/}},
and the corresponding subroutines in the library package
S-lang/ISIS Timing Analysis Routines
\footnote{{\tt http://space.mit.edu/cxc/analysis/SITAR/}}.
Having computed the {\it changepoints} of a feature,
we take the wavelength center as an input for our
{\tt CONTINUUM + LINES} global model. The statistical significance
is afterwards measured as the distance from the continuum
to the core of the line,
in units of the Poissonian error $\sigma_p$.

\section{Fits to the SSS spectrum}
\label{fitssss}

\begin{figure}
\resizebox{12cm}{!}{
\includegraphics[angle=-90]{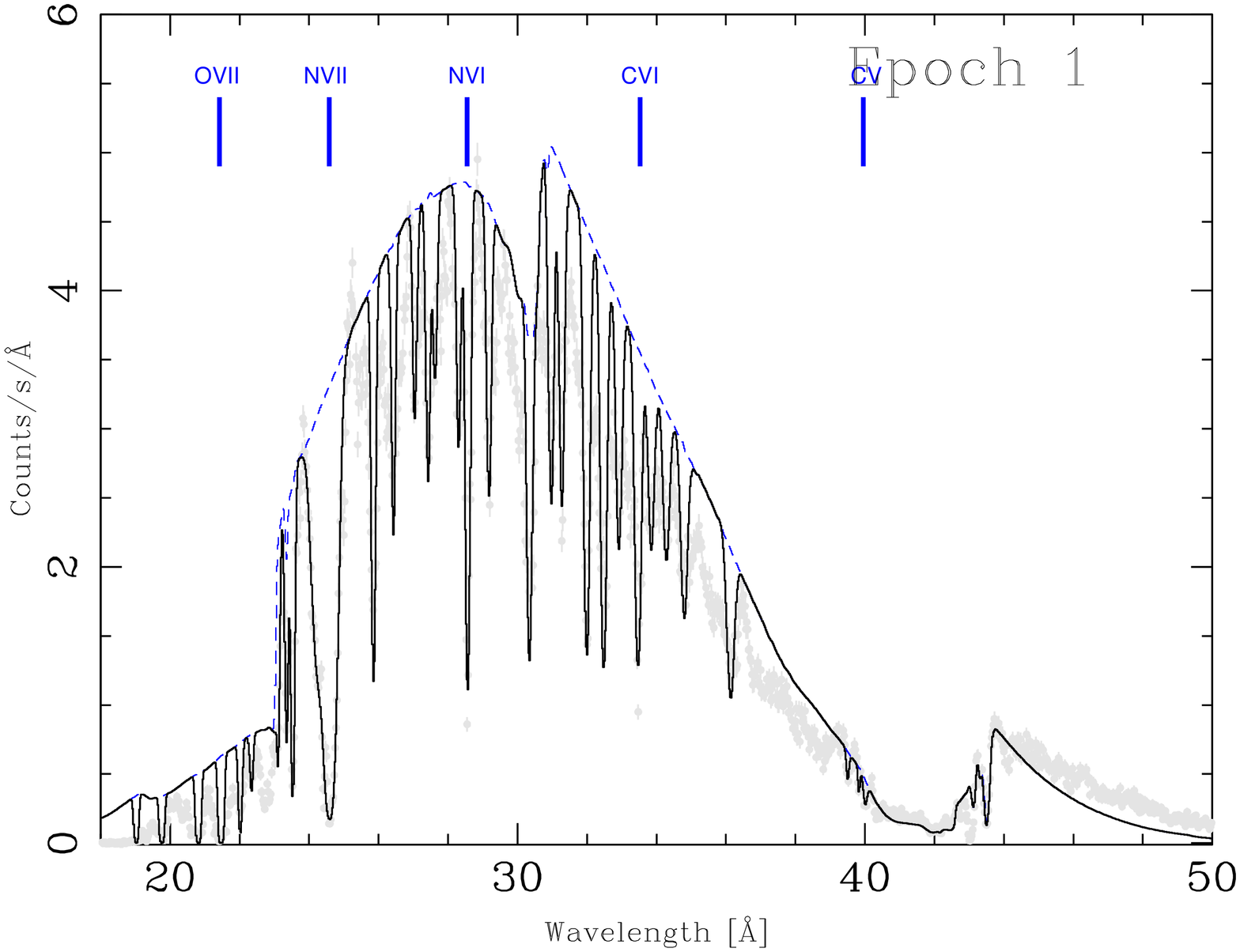}
\includegraphics[angle=-90]{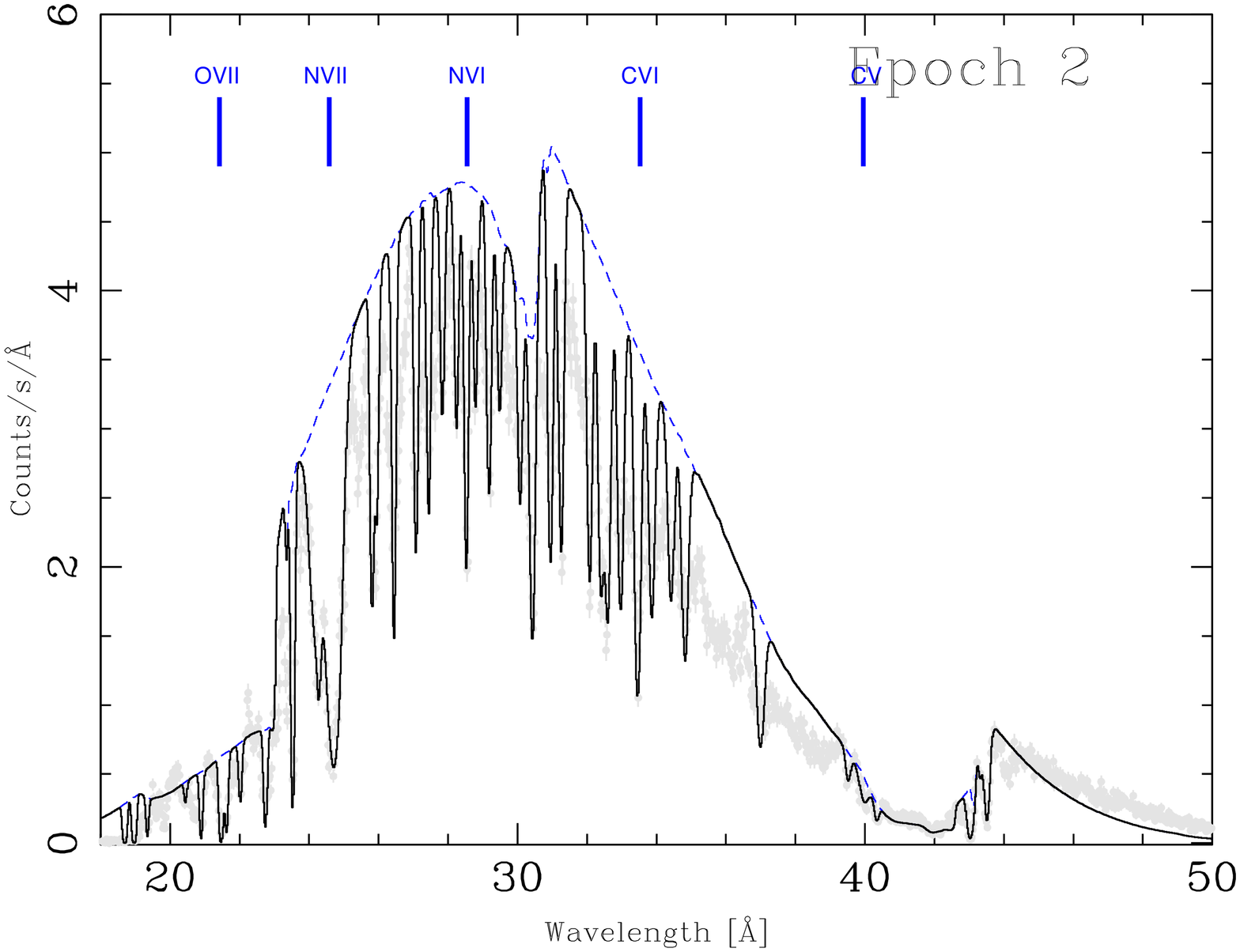}
\includegraphics[angle=-90]{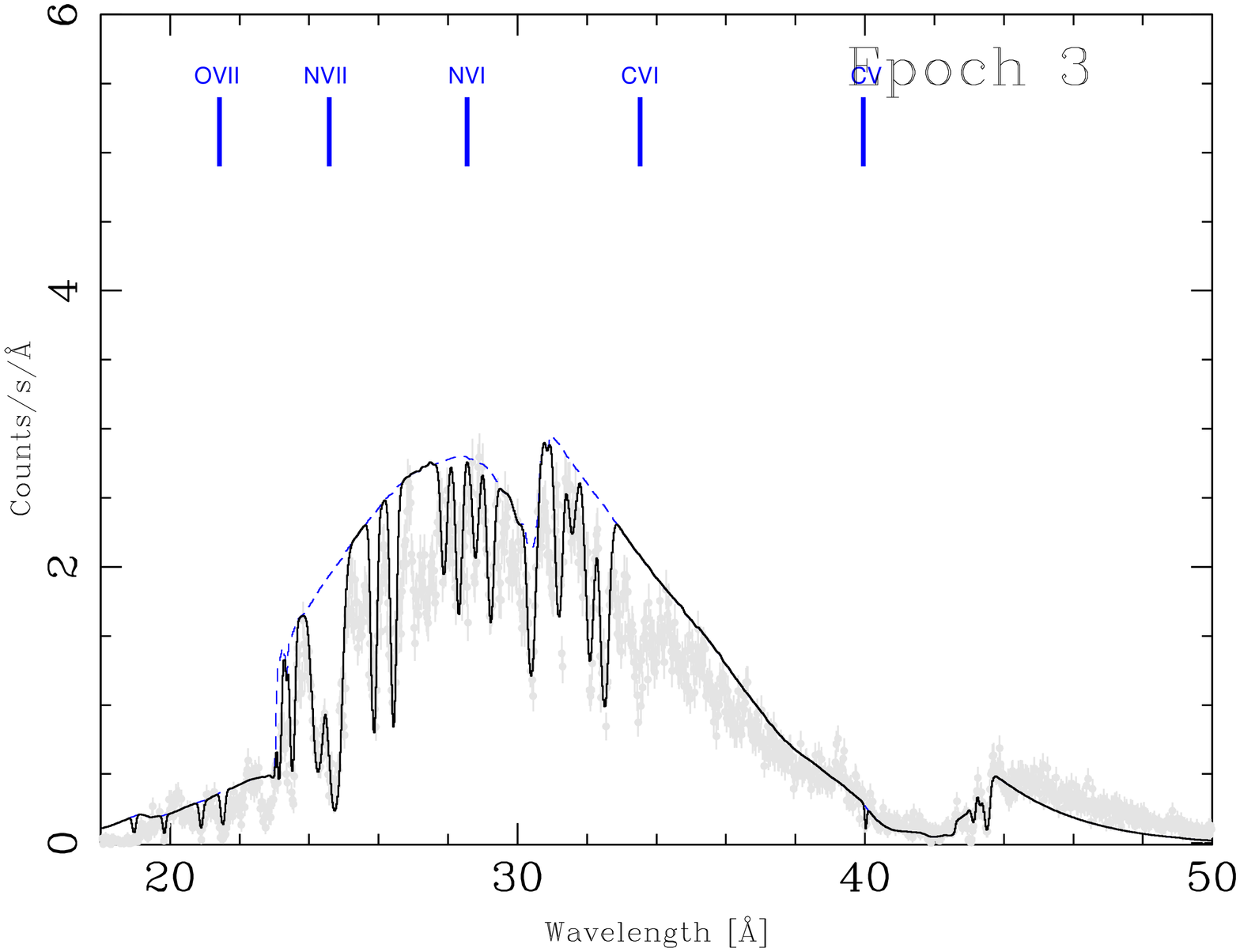}
}
\caption{
Chandra LETGS $\pm$ 1 order spectrum of \sgr, in the soft X-rays bandpass, for epochs (1,2,3).
It is shown the black-body continuum as the thick solid line.
The dashed line, is the global {\tt CONTINUUM + LINES} model we use, in order to characterize the absorption of the system.
\label{spe1}}
\end{figure}

Figure \ref{spe1}
shows the
broad-band
X-ray
spectra
of V4743 Sgr,
by epoch, with the continuum described in \S \ref{cont1}
overplotted as a thick blue line.
This continuum is then modified by multiplicative Gaussians:
\begin{equation}
M(E)=
\exp \left( \tau \times \exp\left[-\frac{(E-E_0)^2}{(2\sigma^2)}\right] \right),
\end{equation}
with
\begin{equation}
\tau = \frac{\tau_0}{\sqrt{(2\pi)}\sigma}.
\end{equation}
Where $\tau_0$ is the optical depth at the core of the line.
The width is given by $\sigma$. And the energy at the core is
$E_0$.
The number of Gaussians to be used is given by the detection procedure
described in \S \ref{cont1}, with $\sigma_p \geq 2$.
The measurements of the line parameters along with their 
identifications are given in Table
\ref{tbl1}, corresponding to epoch 1 
(other Tables are not shown for lack of space).

In the first
column we have the observed
center of the line, in the second column the 
statistical significance of the detection.
In the third column, the candidate ionic specie for that line.
In the fourth, the theoretical wavelength of the transition.
In the fith, the atomic transition of the line.
In the sixth, the Doppler shift of the core of the line.
In the seventh, the 
equivalent width of the line (EW)
in milli-angstroms, computed with:
\begin{equation}
{\rm EW}=\int
\left[ 1-\frac{F_{\rm g}(E)}{F_{\rm c}(E)} \right] dE,
\label{ew1}
\end{equation}
where $F_{\rm g}(E)=
F_{\rm c}\times M(E)$,
$F_{\rm c}(E)$ is the continuum at the energy $E$.
The ion column densities ($N_j$)
are derived from the identifications in the Tables,
assuming that the equivalent widths lie on the linear part of the curve of growth,
and are quoted in colunm 8. 
That is we calculate $N_j$, from:
\begin{equation}
\frac{EW_{\lambda}}{\lambda}=
\frac{\pi e^2}{m_e c^2} N_j\lambda f_{ij},
\label{cd1}
\end{equation}
where
$m_e$ is the electron mass, $c$ the speed of light,
and $f_{ij}$, the oscillator strength of the transition
between levels, $i$ and $j$.


\section{Kinematic modeling of the pre-decay phase}
\label{physmod}

The SSS spectrum originates from the extended atmosphere of
the white dwarf. This is a photoionized 
optically thick
expanding wind (\cite{vanrossumness09}). Here we start by constructing
a kinematic model for such a wind, based on the observed spectra, and we use
this model to extract various important physical parameters from the observations.

\begin{figure}
\resizebox{12cm}{!}{
\includegraphics[angle=-90]{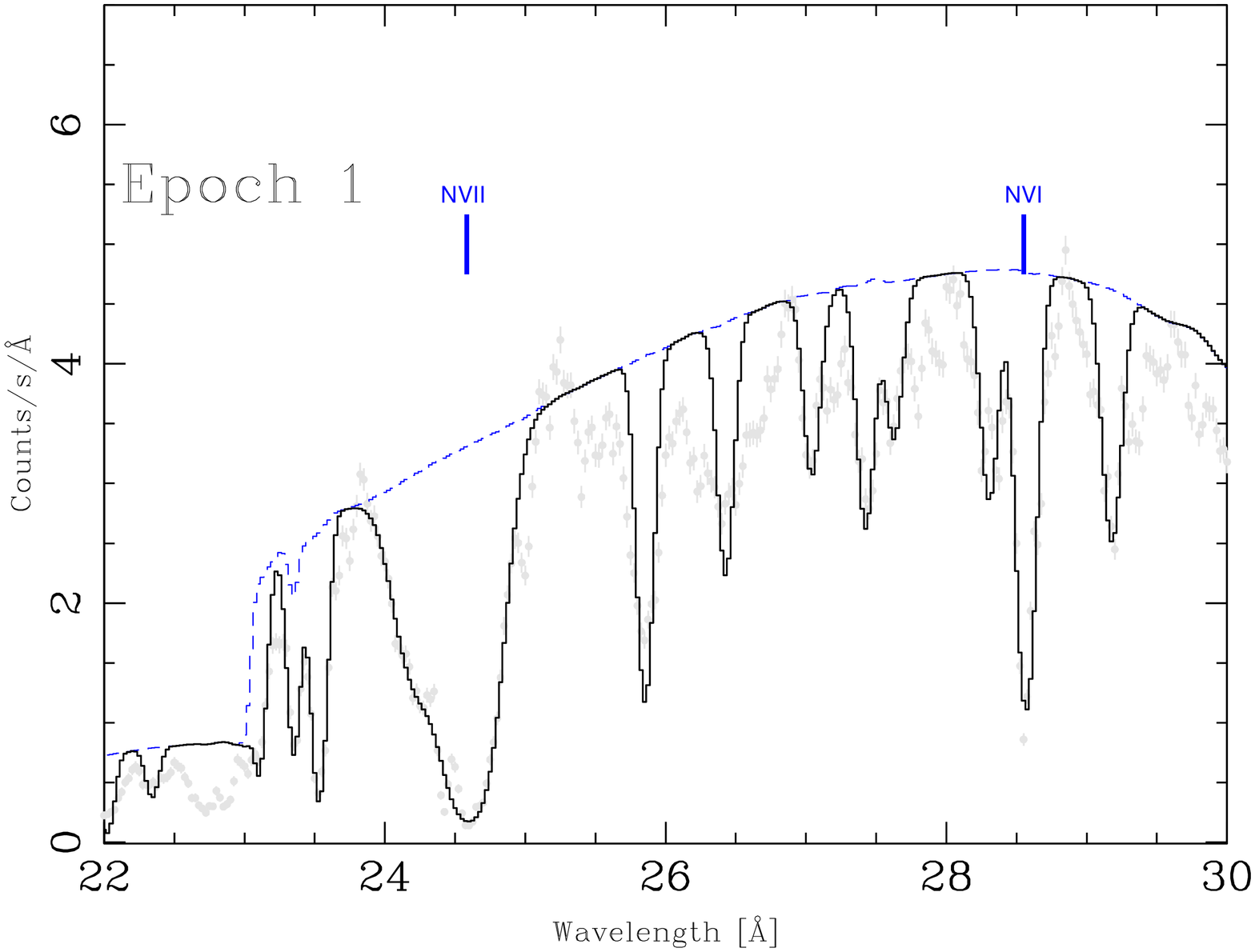}
\includegraphics[angle=-90]{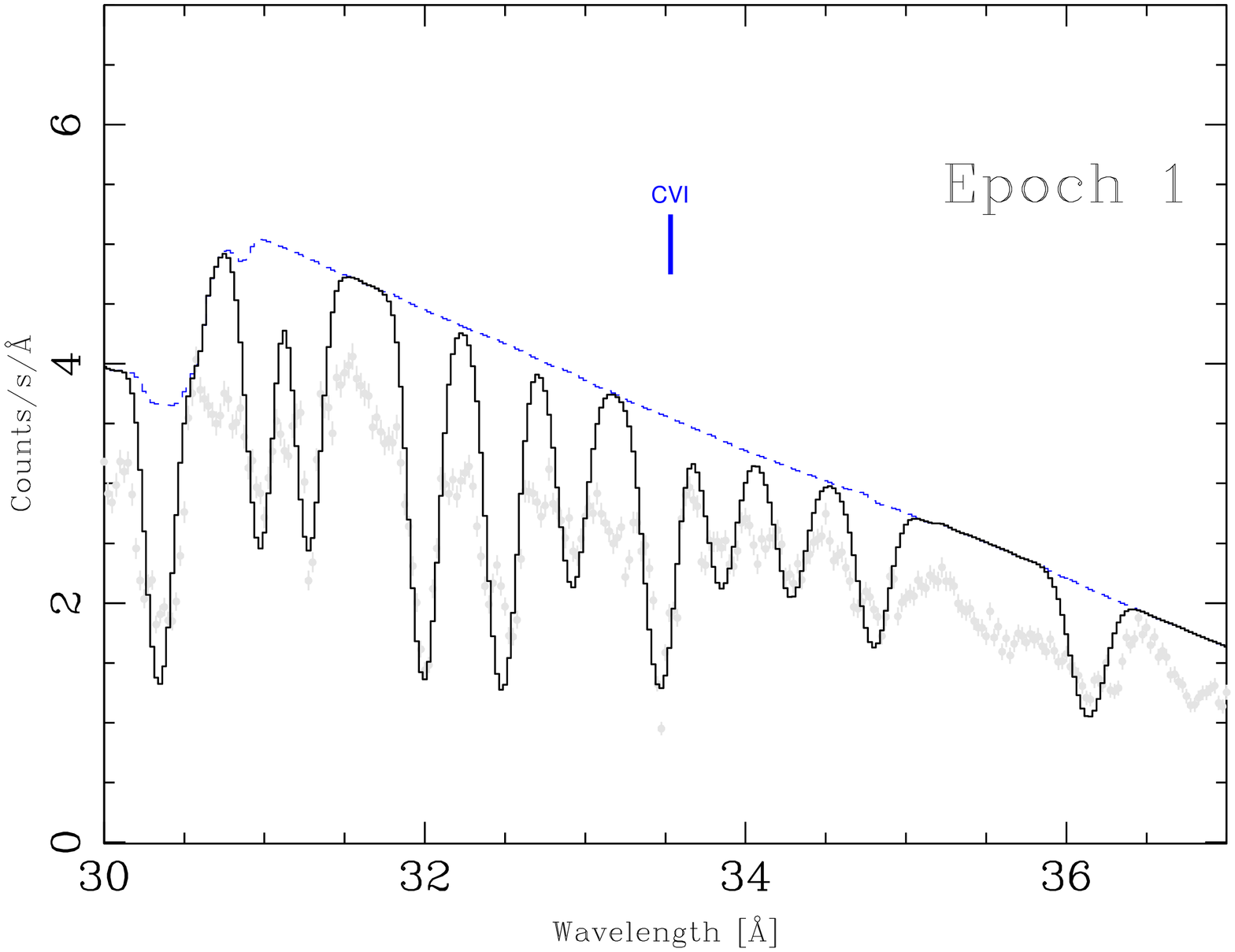}
}
\caption{
{\it Chandra} LETGS $\pm$ 1 order spectrum of Sgr. V4723, in the range 22-37 \AA, for epoch 1.
Several absorption features are identified and modeled with multiplicative Gaussians (solid line) modifying a blackbody continuum (dashed line).
\label{spe4}}
\end{figure}

We assume spherically concentric
shells photoionized
by
a simple blackbody
with integrated
$1-1000$ Ryd luminosity
$L=10^{37}$ \ergs,
and temperature
of
$kT=37$ eV,
which form
an expanding envelope.
At this point, we do not solve the equations of radiative transfer
between spherical clouds,
but assume that there is a relation between radial location
and velocity field, and the ionizing flux dilute only geometrically
as $\propto r^{-2}$.
The gas composition is made of
H, He, C, N, O, Ne, Mg, Si, S,
Ar, Ca and Fe.
We use the abundances of \cite{grevesse1996a}
in our models (and use the term {\it solar} for these abundances).
From the continuity equation for a constant mass loss rate, we adopt a density profile for the expanding envelope
as described by a beta-law,

\begin{equation}
w(x)=w_0+(1-w_0)\left(1-\frac{1}{x}\right)^\beta,
\label{vel1}
\end{equation}
where $w$ is the velocity 
of the wind normalized to the terminal velocity $v_{\infty}$, $w_0$ is the
normalized velocity at the base of the wind and $x$ is the distance
normalized to the radius of the central core $x=r/r_0$.
The parameter $\beta$ is the quantity governing
the slope of the velocity with the distance.
The other function responsible for the variation of the optical depth with
velocity is the ion density. 
This is given by
\begin{equation}
n_i(x)=A_E\times n_H(x)\times
q_i,
\end{equation}
where $A_E=n_E/n_H$ is the abundance
of the element with respect to H, $q_i=n_i/n_E$
is the ionization fraction, and
$n_H(x)$ is
the gas number density
\begin{equation}
n_H(x)=n_0 x^{-2} w^{-1}=\left(\frac{\dot{M}}{4\pi \mu m_H}
r_0^{-2} v_{\infty}^{-1}\right) x^{-2} w^{-1}
\label{nh1}
\end{equation}
where $\dot{M}$ is the mass loss rate and
$\mu m_H$ is the average mass of the
particles.
At this point we need to adopt a model for the ionization balance.
We assume that
we can compute
the ionization fraction $q_i$ of each specie
as a function
of
an ionization
parameter $\xi$
(the ratio of
the ionizing flux $F$ to the gas density $n_H$)
at each radial point.
$\xi$ for the optically thin case is
\begin{equation}
\xi=\frac{4\pi F}{n_H}=\frac{L_{ion}}{r^2 n_H},
\label{xi}
\end{equation}
where $L_{ion}$ is the ionizing luminosity of the source,
but if the space between the source and the photoionized region is
optically thick 
\begin{equation}
\xi=\frac{4\pi F}{n_H}=\frac{L_{ion}\times \exp[-\tau(r)]}{r^2 n_H},
\label{xithick}
\end{equation}
where $\tau(r)$ is the optical depth up to the observed gas.
By combining equations
(\ref{nh1}) and (\ref{xi})
\begin{equation}
\xi=\xi_0\times \exp[-\tau(r)] w,
\label{xiw}
\end{equation}
where $\xi_0=L_{ion}/(r_0 ^2 n_0)$.
This is what we call, a spherical wind,
because the density profile is that of
a spherically symmetric shell.
Due to additional complexities that can be associated
to the expansion of the nova and to take into
account deformation of the density profile
(due to magnetic fields for instance),
we re-write the density as:

\begin{equation}
n_H(x)=n_0 x^{-2+\kappa} w^{-1},
\label{nh}
\end{equation}
where a positive value of $\kappa$ implies that the
gas flow dilutes more slowly than in a free
spherical expansion, i.e. that there are sources of
gas embedded in the flow, or that the flow is confined.
A negative value corresponds to sinks of gas in the flow,
or expansion of an initially confined flow in a flaring
geometry.
Here and in what follows,
we set $w_0=0$ for simplicity.
We then find that
\begin{equation}
\xi=\xi_0\times \exp[-\tau(r)] w\times
(1-w^{1/\beta})^\kappa,
\label{xiwp}
\end{equation}
and this is the relationship
between the ionization parameter and
the radial velocity for
a non-spherical, optically thick wind.
The two
panels in Figure \ref{spe4}
shows details of the spectrum and the results of our model. Most of the Gaussian features match 
well the
observed spectrum, more specifically the centroids of the features are well located,
a good element for our purposes. However, the fit is far from being statistically acceptable.

\section{Ionizing Luminosity, Mass Loss Rate, and Kinetic Luminosity of the Flow}
\label{moi-results}

We seek to constrain key physical parameters of the wind such as its mass loss rate and 
kinetic luminosity from the observed spectra and the relations derived above 
(see Table \ref{tbl5} for a summary).
But, first one needs to answer a basic question, 
is the wind optically thin or optically thick?

\begin{figure}
\begin{center}
\resizebox{6cm}{!}{\includegraphics[angle=-90]{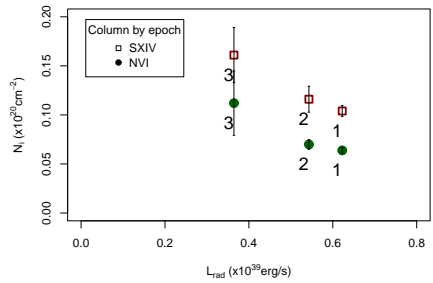}}
\caption{
Column densities
of S~{\sc xvi} and N~{\sc vi} resonant lines vs.  apparent luminosity
for the three epochs of observation.
These two column densities are easily measured in the spectra and
span a large range in ionization.
\label{colvel1}}
\end{center}
\end{figure}

To answer this we look for possible changes in the ionization of the spectra 
among the three epochs of observation. 
In Figure \ref{colvel1} we look at the relative column densities 
of S~{\sc xvi} and N~{\sc vi} resonant lines vs.  apparent luminosity 
for the three epochs of observation.

These two column densities are easily measured in the spectra and 
expand a large range in ionization. The ratio of column densities 
is notoriously flat, and so are the ionization of the spectra, 
while the apparent luminosity changes by a factor of $\approx 2$. 
This means that the three spectra steam from different radial distance, 
being the region farthest from the source in epoch 1 when the luminosity 
is maximum and nearest in epoch 3. It also means that if $\xi$ is to remain 
constant for varying $L_{rad}$ and the wind is spherical it cannot be optically thin, 
but thick (see equation~\ref{xi}).

\begin{table*}
\caption{Parameters of the system: Mass loss rate $\dot{M}$, $\dot{M}/L_{\rm ion}$ and $L_{kin}/L_{\rm rad}$.}
\label{tbl5}
\begin{tabular}{c c c c}
\hline \hline
Epoch &
$\dot{M}$$^{\rm a}$ &
$\dot{M}/L_{\rm ion}$$^{\rm b}$ &
$L_{kin}/L_{\rm rad}$

\\
\hline
I (CVI) & 4.2 & 2.63 & 3.02587 \\
I (NVI) & 5.3 & 3.32 & 3.52105 \\
I (OVII) & 2.5 & 1.55 & 1.07782 \\
II (CVI) & 4.3 & 2.71 & 3.34455 \\
II (NVI) & 6.2 & 3.92 & 5.77883 \\
II (OVII) & -- & -- & -- \\
III (CVI) & -- & -- & -- \\
III (NVI) & 0.1 & 0.09 & 0.00007 \\
III (OVII) & 1.5 & 0.93 & 0.23078 \\
\hline
\end{tabular}

(a) In units of $\times 10^{-4} ~ (\frac{L}{L_{38}}) ~ M_{\sun}/yr$.
(b) In units of $\times 10^{-16} ~ {\rm gr}/{\rm ergs}$.

\end{table*}

\begin{figure}
\begin{center}
\resizebox{6cm}{!}{\includegraphics[angle=-90]{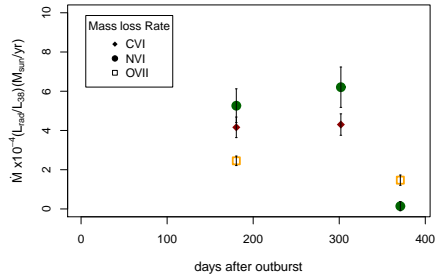}}
\caption{
Mass loss rate during days 180.4, 301 and 371
after outburst. We use three spectral absorption lines to compute this
quantity:
C~{\sc vi $\lambda$33.734}, 
N~{\sc vi $\lambda$28.787} and
O~{\sc vii $\lambda$21.602}.
Because of the luminosity is model-dependent, $ \dot{M} $
is give as a function of it.
$L_{38} $ is luminosity in units of $10^{38} $ \ergs.
\label{phys8}}
\end{center}
\end{figure}

From equations \ref{nh1} and \ref{xithick} one can write the mass loss rate as
\begin{equation}
\dot{M}=4\pi \mu m_H L_{\rm ion} \times \exp(-\tau) \left(\frac {v}{\xi}\right)
\label{massloss2}
\end{equation}
where $ \mu m_H$ is the average mass of the particles.
The ratio $({v}/{\xi})$ can be independently determined from observations for each of the species observed in the spectra. 
The values of $v$ were measured from the centroids of the lines;
C~{\sc vi $\lambda$33.734},
N~{\sc vi $\lambda$28.787} and
O~{\sc vii $\lambda$21.602}
for the three epochs under consideration (see 
Tables 3~-~5). The values of $\xi$ are taken from the 
ionization balance curves of \cite{kallman2001a}. 
Figure \ref{phys8} presents the results for $\dot{M}$ as derived from different
species and on different epochs. The spread in values from different
species is probably due to variations in the optical depth up to the regions 
where different ions form. The optical depth $\tau$ is a complex 
function that depends on the column density of gas and its physical 
conditions up to the region responsible for the observed absorption troughs. 
The overall results are for an upper limit to  
$\dot{M} \approx 4\times 10^{-4} ~ (\frac{L}{L_{38}}) ~ M_{\sun}/yr$. 

\begin{figure}
\begin{center}
\resizebox{6cm}{!}{\includegraphics[angle=-90]{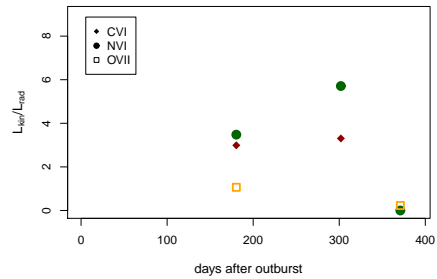}}
\caption{
Efficiency of the amount of energy used for the motion of the ejecta,
measured as the ratio between the kinetic luminosity $L_{\rm kin}$ and the radiated luminosity
$L_{\rm rad}$. 
We use three spectral absorption lines to compute this
quantity:
C~{\sc vi $\lambda$33.734},
N~{\sc vi $\lambda$28.787} and
O~{\sc vii $\lambda$21.602}.
The three lines are consistent with an efficiency of the order of one.
See text for details.
\label{phys10}}
\end{center}
\end{figure}

The kinetic energy of the flow is defined as
\begin{equation}
L_{kin}=\frac{1}{2} \dot{M} v^2.
\end{equation}
Our results for the ratio of $(L_{kin}/L_{rad})\times \exp{\tau}$ in epochs 1~-~3 are depicted in Figure \ref{phys10}.
The numeric values in this figure can be regarded as upper limits to the ratio of kinetic to radiative energy in the wind.
Interestingly, this ratio is less than one
in epoch 3, suggesting that the flow has fallen out of equipartition.
This can be understood if by this time in the expansion radiative pressure has become insufficient to overcome the gravitacional
potential. If this result is applicable to other novae it means that ejecta mass determinations based on the assumption of equipartition are only reliable when 
near the spectral peak, which is also the kinetic peak. During other phases, the
mass outflow is very small.

\section{Analysis and discussion}
\label{discu1}

Here we present an analysis
and following of a discusion of the:
(1) early hard spectrum,
(2) sss-phase and
(3) the post-sss
phase.

\subsection{Analysis of early hard spectrum}
\label{m:early}

V4743\,Sgr is not the first nova for which a hard
X-ray spectrum was found before the WD has become visible
in X-rays. Spectral model fits to such spectra imply
that the X-ray emission originates from an optically
thin collisional plasma, leading to
the interpretation of shock fronts within the ejecta
(\cite{obrien94}).

For this reason, we chose a model that produces an
emission line spectrum arising in an optically thin
plasma with an isothermal electron temperature, $T_e$,
in order to model the ACIS-S spectrum taken on day
50.2. We have used {\tt xspec}
\footnote{{\tt http://heasarc.nasa.gov/docs/xanadu/xspec/manual}}
to fit the model to the data while
accounting for all instrumental effects. The vapec model
is based on atomic data calculated by the Astrophysical
Plasma Emission Code (APEC, v1.3:, \cite{smith01})
and allows the abundances of He, C, N, O, Ne, Mg, Al, Si, S,
Ar, Ca, Fe, and Ni to vary. The associated database, APED, is a
more recent compilation of the atomic data contained in the
MEKAL database which was used by \cite{balm98} for the early
emission of V1974\,Cyg. The underlying assumptions are the same,
i.e., the plasma is optically-thin and in collisional equilibrium.
The principal parameters are temperature, normalization, and abundances
relative to Solar. Since the plasma is unlikely to
be isothermal, the sum of several APEC models can be used as an
approximation of a continuous distribution of temperatures. 

We have corrected for interstellar absorption
assuming a fixed value of $N_{\rm H}=1.4\times10^{21}$\,cm$^{-2}$
\cite{lyke}), using the absorption model {\tt tbabs}.
The normalization parameter can be converted to volume emission
measure ($VEM$, units cm$^{-3}$), assuming a distance of
3.9\,kpc. We found isothermal models statistically unsatisfactory
with none achieving a reduced $\chi2$ better than $\chi2_{\rm red}=1.70$
(66 degrees of freedom, $dof=N-n_p$). With 2-temperature (2-$T$) models,
we found significantly better fits (see Table~\ref{models}).
For the best-fit model, we found an F-test value of 64
and a probability of $1.8\times 10^{-6}$, thus the second model
component is statistically justified. We found no further improvement
with 3-$T$ models.

We first assumed solar abundances, scaled with a
single factor for all abundances, but we found poor fits. Next
we iterated the abundances of N, O, Ne, Mg, Si, S, and
Fe, with other elements fixed at their solar values because no
strong lines from these species are observable. The values of
the abundances are poorly constrained. For Fe we find only an
upper limit of $<0.04$. We caution, however, that the low Fe
abundance may be an artifact from approximating a continuous
temperature distribution with only two isothermal components.
Likewise, the abundances of the other elements are poorly
constrained, and no robust conclusions can be drawn from the
best-fit values.
The model parameters are listed in Table~\ref{models}.

\begin{table}
\begin{flushleft}
\renewcommand{\arraystretch}{1.1}
\caption{\label{models}Models to ACIS spectrum (50.2 days after discovery)}
{
\begin{center}
\begin{tabular}{llr}
\hline
Param. & Unit & Value$^a$\\

k$T_1$ & keV\dotfill & $0.25-0.35$ \\
$\log(VEM_1)^c$ & cm$^{-3}$\dotfill & $56.97-57.35$ \\
k$T_2$ & keV\dotfill & $2.7-15$ \\
$\log(VEM_2)^c$ & cm$^{-3}$\dotfill & $56.05-56.28$ \\
$N_{\rm H}$ & cm$^{-2}$\dotfill & $(1.4\times 10^{21})^b$ \\
flux$^d$ &$10^{-12}$\,erg\,cm$^{-2}$\,s$^{-1}$\dotfill & $1.2-2.4$ \\
$L_X^{c,d}$ & $10^{32}$\,erg\,s$^{-1}$\dotfill & $1.7-3.5$\\
$\chi2_{\rm red}$ ($dof$) &\dotfill & 1.16 (64)\\
\hline
\end{tabular}
\end{center}
}
$^a$90\% uncertainty ranges\\
$^b$fixed\\
$^c$assuming distance 3.9\,kpc\\
$^d0.2-10$\,keV\\
Abundances of N, O, Ne, Mg, Si, S, and Fe were varied (see text)
\renewcommand{\arraystretch}{1}
\end{flushleft}
\end{table}

\subsection{Discussion of early hard spectrum}
\label{disc:early}

The optically thin models used for fitting the early hard
spectrum have originally been developed
for the X-ray spectra of the Solar and stellar coronae. Our spectrum
differs only in the X-ray luminosity and the likely different chemical
composition of nova ejecta. The possibility could be considered that
the early hard emission is of a similar origin. The
Solar corona is powered by magnetic fields that are generated
in the tachocline, the sheer layers between the inner radiation
zone and the outer convection zone. In a nova, magnetic fields could
be generated by a similar dynamo, located in the interface
between the ejected envelope and the WD surface that is rigidly
rotating underneath the ejecta. If V4743\,Sgr is an intermediate
polar as suggested by \cite{kang06}, and \cite{dobrness09}, 
the permanent magnetic field of the WD could amplify
dynamo-generated magnetic fields that power a corona.
However, these considerations are solely
based on plausibility arguments and are difficult to test, as
no coherent models for the production of coronal emission
exist. While the influence of magnetic fields can not necessarily
be discarded, magnetic fields are not part of the models assumptions
of the optically thin thermal models. Other possibilities can
thus be considered.

More commonly accepted is the idea of interpreting the
collisional nature as originating from a shock-heated
plasma, although it is not clear whether the shocks originate
from within the ejecta or from interactions with circumstellar
material or the stellar wind of the companion.

Shocks with the stellar wind of the companion can only
produce X-ray emission that is strong enough to be detectable
at the given large distance if the stellar wind is
sufficiently thick, and that requires a symbiotic nova such
as RS\,Oph (\cite{rsophshock}). In those systems,
the companion is an evolved giant or subgiant.
The ejecta run into this medium and
dissipate some of their kinetic energy in the form of X-ray
emission in the resulting shock (\cite{obrien92}). In Classical
Novae like V4743\,Sgr, no strong stellar wind is present and no
hard X-ray emission is expected from this production mechanism.
One indication that V4743\,Sgr
is not a symbiotic nova is its short orbital
period of 6.7 hours (\cite{kang06}), which is much shorter than
typical orbital periods of symbiotic novae of several hundred days.
We also consider shock interactions with circumstellar material
unlikely, since the required density is higher than models of
nova binary systems would suggest (\cite{lloyd92}). We have no
reason to assume a higher density of the ambient medium as believed
in GK\,Per (\cite{bode04}) or V458\,Vul (\cite{v458}),
where a planetary nebula was found to surround the nova.

\cite{obrien94} developed a shocked-gas model for the specific
case of V838\,Her and argued that the shock-heating must take
place as a result of the interaction of different components
{\em within} the ejecta. This ``interacting winds'' model
has been refined by \cite{lloyd95}. The complexity of the
absorption lines that we found in the later SSS spectra indicates
that the ejecta are not homogeneous, which is an important
ingredient for this model. The presence of different
velocity components could aid the development of shocks
within the ejecta.

On the other hand, in order to produce enough X-ray emission
in a shock, sufficiently high densities are needed, which
unavoidably also have a high opacity. It has not been tested
whether X-ray resonance lines of H-like ions can escape such
a plasma. Furthermore, at least two distinct episodes of ejection
are needed, providing slow-moving ejecta from an earlier
outburst and fast-moving ejecta from a later ejection event.
While this has been observed in V2362\,Cyg,
no explanation on how such secondary events could occur have
been found. It could be possible, on the other hand, that
some material is falling back, colliding with the expanding
ejecta.

With all the given arguments, the cause for the early hard
emission can not be identified with certainty. At this time, no
deep X-ray spectrum of the early hard emission phase of a nova
has been taken that could be used to pose constraints on the
different causes.

\subsection{Analysis of the SSS phase}
\label{disc:sss}

The details of the grating spectra confront us with a high
degree of complexity. The continuum has the shape of a
stellar atmosphere in all observations between days 180.4
and 370, but the absorption lines are blue-shifted and highly
structured. The blue-shifts of all photospheric lines
indicate that expansion is still continuing during
the SSS phase, bringing about (as we will show later) a significant 
mass loss
(\cite{ness09}).
The line profiles
are extremely complex, and not all lines have
the same profile. 
Globally, up to three different velocity-bands
can be observed:
$v_1 \sim-1000$ \kms, $v_2 \sim-2500$ \kms \sp and
$v_3 \sim-6000$ \kms,
coming from ionic species spanning a wide range
of ionization states, from
C~{\sc v} to S~{\sc xiv}
(see Table \ref{tbl1}).

\begin{table*}
\caption{Line properties of V4743 Sgr., LETGS epoch 1 (day 180.39).}
\label{tbl1}
\begin{tabular}{c c c c c c c c c}
\hline \hline
$\lambda _{obs} $$^{a}$ &
$\sigma _p $$^{b}$ &
Ion $^{c}$ &
$\lambda _0$ $^{d}$ &
Atomic transition &
Velocity (\kms) &
EW (m\AA) &
N$_i$ $^{e}$ &
flag
\\
\hline
$ 19.768 \pm 0.003 $ & 8 & N~{\sc vii } & 19.826 & 1s $^2$S - 4p $^2$P & $ -873 \pm 51 $ & $ 88 \pm 10 $ & $ 0.52 \pm 0.06 $ & i\\
$ 20.818 \pm 0.003 $ & 10 & N~{\sc vii } & 20.910 & 1s $^2$S - 3p $^2$P & $ -1312 \pm 45 $ & $ 98 \pm 13 $ & $ 0.19 \pm 0.03 $ & i\\
$ 21.470 \pm 0.001 $ & 11 & O~{\sc vii } & 21.602 & 1s$^2$ $^1$S - 1s 2p $^1$P & $ -1832 \pm 16 $ & $ 104 \pm 1 $ & $0.03\pm0.01 $ & s\\
$ 22.021 \pm 0.002 $ & 11 & O~{\sc vi } & 22.020 & 1s$^2$ 2s $^2$S - 1s 2s(3P) 2p $^2$P & $ 11 \pm 25 $ &$215\pm31 $ & $0.05\pm0.01 $ & i\\
$ 22.345 \pm 0.004 $ & 12 & O~{\sc v } & 22.360 & 1s$^2$ 2s$^2$ $^1$S - 1s2p $^1$P & $ -334 \pm 47 $ &$281 \pm 76 $ & $ NA $ & i\\
$ 23.106 \pm 0.001 $ & 20 & N~{\sc vi } & 23.277 & 1s$^2$ $^1$S - 1s 5p $^1$P & $ -2202 \pm 19 $ & $ 271 \pm 30 $ & $ 2.00 \pm 0.22 $ & i\\
$ 23.355 \pm 0.002 $ & 17 & O~{\sc i } & 23.450 & 1s$^2$ 2s$^2$ 2p$^4$ - 1s 2s$^2$ 2p$^5$ & $ -1215 \pm 20 $ & $ 291 \pm 32 $ & $ 0.42 \pm 0.05 $ & g\\
$ 23.534 \pm 0.001 $ & 22 & N~{\sc vi } & 23.771 & 1s$^2$ $^1$S - 1s 4p $^1$P & $ -2989 \pm 13 $ & $ 257 \pm 18 $ & $ 0.88 \pm 0.06 $ & i\\
$ 24.604 \pm 0.002 $ & 24 & N~{\sc vii } & 24.779 & 1s $^2$S - 2p $^2$P & $ -2121 \pm 19 $ & $ 975 $ & $ 0.26 \pm 1.86 $ & s\\
$ 25.856 \pm 0.001 $ & 24 & Ca~{\sc xi } & 26.442 & 2p$^6$ $^1$S-2p$^5$ 4s $^1$P & $ -6644 \pm 13 $ & $ 351 \pm 20 $ & $ 1.50 \pm 0.09 $ & i\\
$ 26.430 \pm 0.002 $ & 20 & Ca~{\sc xi } & 26.962 & 2p$^6$ $^1$S - 2s 2p$^6$3p $^1$P & $ -5915 \pm 20 $ &$407 \pm 35$ & $0.17\pm0.01$& s\\
$ 27.044 \pm 0.003 $ & 21 & Si~{\sc xiii } & 27.341 & 1s 2s $^3$S - 1s 4p $^3$P & $ -3259 \pm 31 $ & $ 452 \pm 57 $ & $ 0.37\pm0.05$ & i\\
$ 27.429 \pm 0.002 $ & 22 & Ar~{\sc xiv } & 27.464 & 2s$^2$ 2p $^2$P - 2s$^2$ 3d $^2$D & $ -382 \pm 22 $ & $ 447 \pm 41 $ & $0.04\pm0.01$ & i\\
$ 27.629 \pm 0.003 $ & 18 & Ar~{\sc xiv } & 27.636 & 2s$^2$ 2p $^2$P - 2s$^2$ 3d $^2$D & $ -76 \pm 36 $ & $ 480 \pm 71 $ & $ 0.23 \pm 0.03$ & i\\
$ 28.309 \pm 0.002 $ & 21 & C~{\sc vi } & 28.465 & 1s $^2$S - 3p $^2$P & $ -1645 \pm 23 $ & $ 484 \pm 46 $ & $ 0.51 \pm 0.05 $ & s\\
$ 28.570 \pm 0.001 $ & 29 & N~{\sc vi } & 28.787 & 1s$^2$ $^1$S - 1s 2p $^1$P & $ -2261 \pm 11 $ & $ 419 \pm 18 $ & $ 0.06 \pm 0.01 $ & s\\
$ 30.346 \pm 0.002 $ & 21 & S~{\sc xiv } & 30.427 & 1s$^2$ 2s $^2$S - 1s$^2$ 3p $^2$P & $ -799 \pm 15 $ & $ 507 \pm 27 $ & $ 0.10 \pm 0.01$ & s\\
$ 30.975 \pm 0.002 $ & 23 & Si~{\sc xii } & 31.012 & 1s$^2$ 2s $^2$S - 1s$^2$ 4p $^2$P & $ -359 \pm 17 $ & $ 558 \pm 33 $ & $ 0.44 \pm 0.03 $ & i\\
$ 31.277 \pm 0.002 $ & 26 & N~{\sc i } & 31.223 & 1s$^2$ 2s$^2$ 2p$^3$ - 1s 2s$^2$ 2p$^4$ & $ 518 \pm 18 $ & $ 572 \pm 35 $ & $ 0.13 \pm 0.01 $ & g\\
$ 31.996 \pm 0.001 $ & 26 & S~{\sc xiii } & 32.239 & 2s$^2$ $^1$S - 2s 3p $^1$P & $ -2261 \pm 13 $ & $ 551 \pm 23 $ & $ 0.10 \pm 0.01 $ & i\\
$ 32.483 \pm 0.001 $ & 24 & Ca~{\sc viii }& 32.770 & 3p $^2$Po - $\infty$ & $ -2626 \pm 12 $ & $ 567 \pm 24 $ & $ 0.21 \pm 0.01 $ & i\\
$ 32.921 \pm 0.002 $ & 19 & Ca~{\sc viii }& 33.120 & 3p $^2$Po - $\infty$ & $ -1801 \pm 22 $ & $ 645 \pm 45 $ & $ NA $ & i\\
$ 33.469 \pm 0.002 $ & 26 & C~{\sc vi } & 33.734 & 1s $^2$S - 2p $^2$P & $ -2357 \pm 16 $ & $ 619 \pm 30 $ & $ 0.09 \pm 0.01 $ & i\\
$ 33.850 \pm 0.003 $ & 17 & S~{\sc xii } & 34.533 & 2s$^2$ 2p $^2$P - 2s 2p(3P) 3p $^2$D & $ -5928 \pm 30 $ & $ 703 \pm 67 $ & $ 0.12 \pm 0.01 $ & i\\
$ 34.286 \pm 0.004 $ & 16 & C~{\sc v } & 34.973 & 1s$^2$ $^1$S - 1s 3p $^1$P & $ -5887 \pm 34 $ & $ 727 \pm 75 $&$0.32\pm0.03$&i\\
$ 34.799 \pm 0.003 $ & 16 & Ar~{\sc ix } & 35.024 & 2p$^6$ $^1$S - 2p$^5$ 4d $^1$P & $ -1926 \pm 28 $ & $ 728 \pm 60 $&$0.10\pm0.01$ &i\\
$ 36.138 \pm 0.003 $ & 15 & S~{\sc xii } & 36.398 & 2s$^2$ 2p $^2$P - 2s$^2$ 3d $^2$D & $ -2142 \pm 27 $ & $ 764\pm57$&$0.04\pm0.01$& i\\
$ 39.508 \pm 0.007 $ & 18 & Mg~{\sc x } & 39.668 & 1s$^2$ 2s $^2$S - 1s$^2$ 5p $^2$P & $ -1212 \pm 54 $ & $ 266 \pm88$&$0.31\pm0.10$&i\\
$ 40.012 \pm 0.008 $ & 8 & C~{\sc v } & 40.268 & 1s$^2$ $^1$S - 1s 2p $^1$P & $ -1905 \pm 58 $ & $ 395 \pm 80 $ & $ 0.03 \pm 0.01 $ & i\\
$ 39.832 \pm 0.006 $ & 9 & Si~{\sc xi } & 40.286 & 2s$^2$ $^1$S - 2p 3p $^3$D & $ -3379 \pm 46 $ & $ 68 \pm 334 $ & $ NA $ & i\\
\hline
\end{tabular}

Errors are 1$\sigma$, computed individually for each of the parameters.
(a) Observed wavelength in \AA.
(b) Statistical significance of the detection.
(c) Probable ion identification.
(d) The latest laboratory wavelength (in \AA) taken from the
National Institute of Standars and Technology ({\tt NIST: http://physics.nist.gov/PhysRefData/ASD/lines\_form.html}).
The K-shell wavelength of the oxygen ions (O~{\sc vi} and O~{\sc v}), are taken from
\cite{garcia2005a}.
(e) Measured column density for the ion in units of $\times 10^{20}$ \cmn.
(f) This is a broad feature, not actually associated to any atomic transition.
Column {\tt flag}: (s) Secure identification
(i) Insecure identification
(vi) Very insecure identification: means that there is no reliable atomic data for this ion.
(g) Galactic line.


\end{table*}

The distinctness
of three velocity components as seen in 
Fig.~\ref{spe4},
suggests that three (or more)
concentric shells with low-density plasma
in between are present. On the other hand, some lines are
much narrower with only one or two components. Clearly, not
all lines originate from the same region, indicating that a
large range of different plasma regions are visible at the
same time.

The amount and duration of mass-loss during the SSS phase can
only be determined from detailed, physically plausible
models. NLTE effects have proven to be crucial
\cite{hartheis97}, and minimum requisite of the model
has to be that it is spherically symmetric and expanding
\cite{vanrossumness09}. See also \S \ref{physmod} for
evidences of this latest assumption.

\begin{figure}
\rotatebox{0}{
\resizebox{10cm}{!}{
\includegraphics{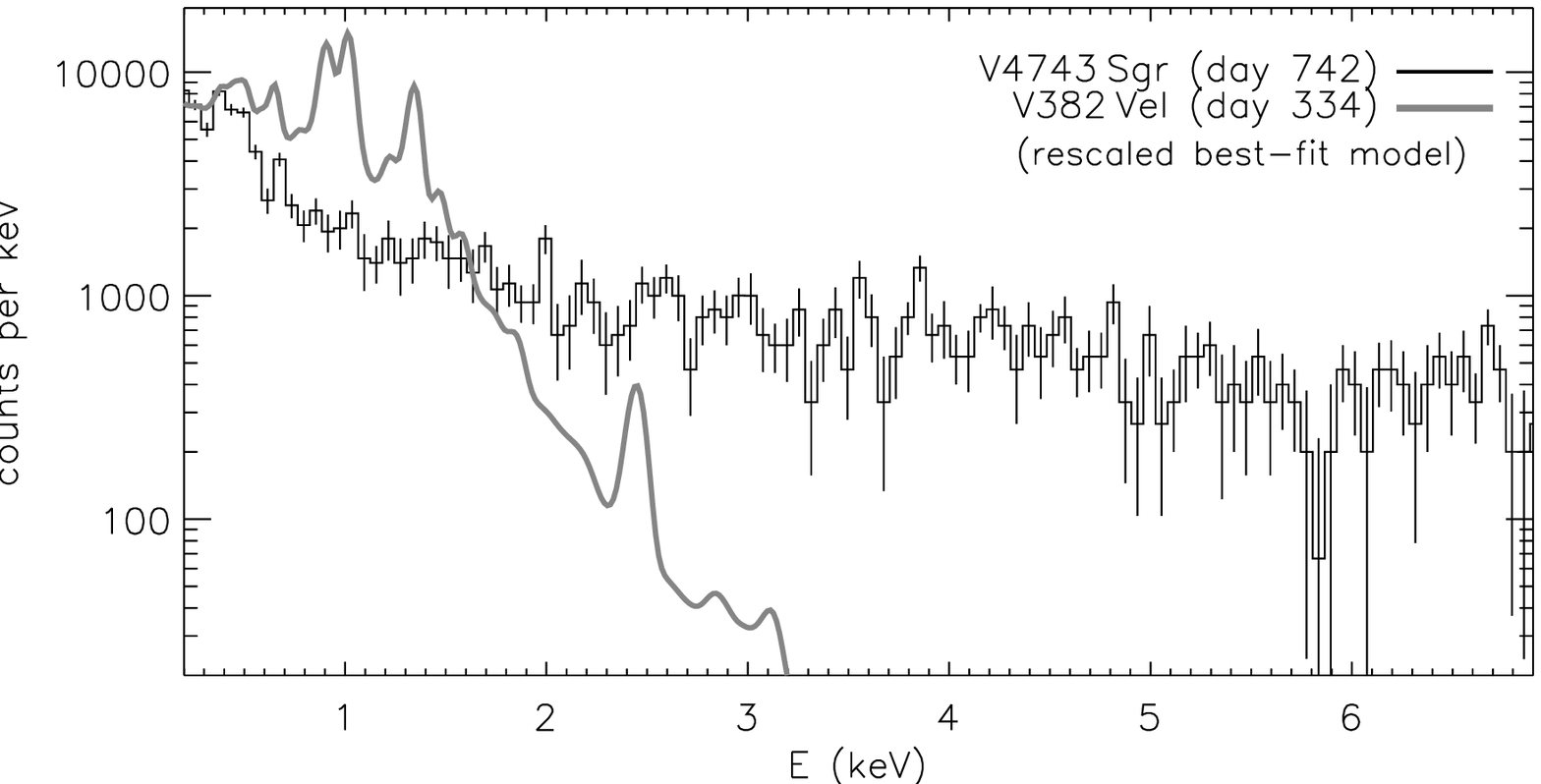}
\includegraphics{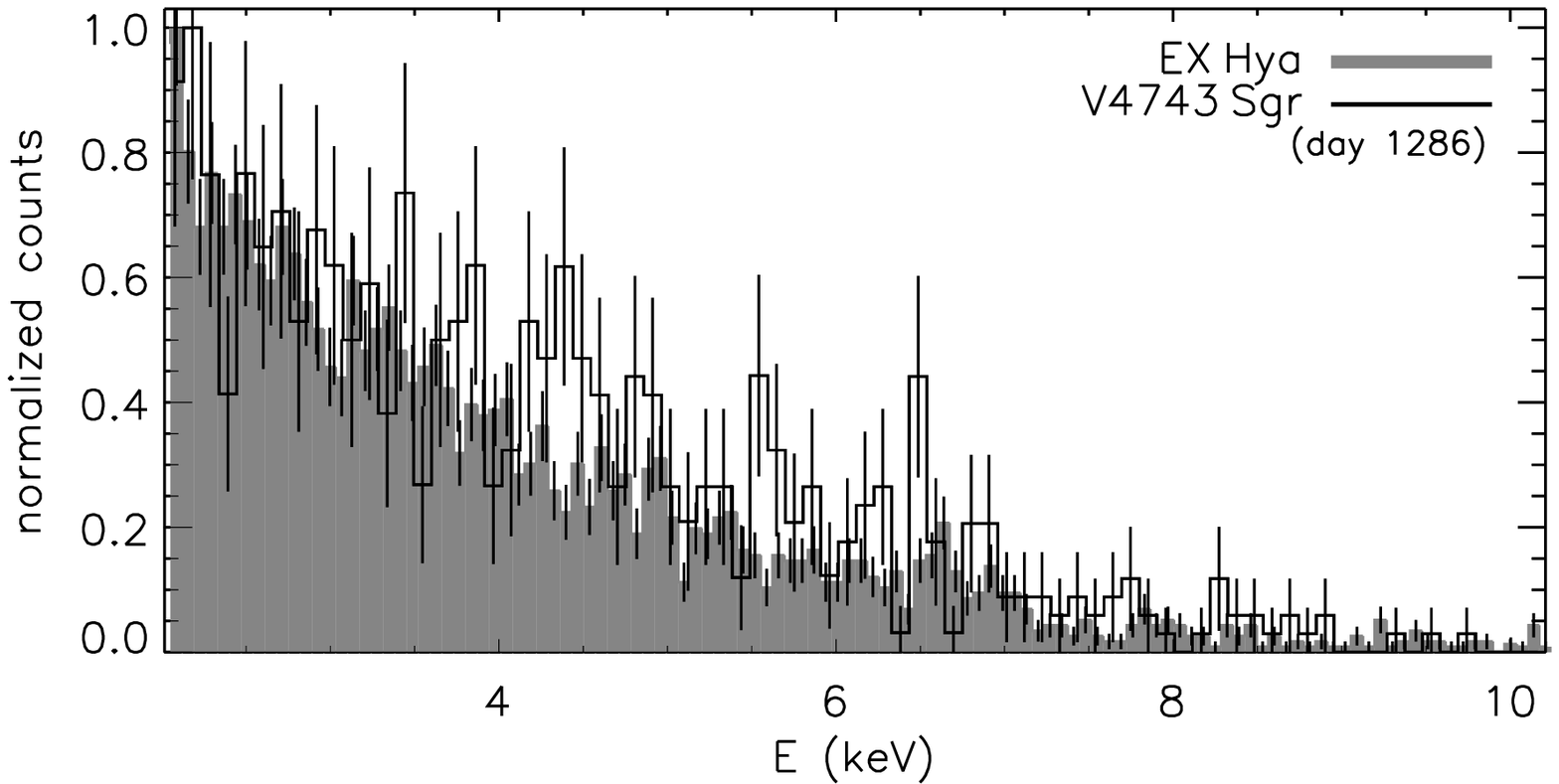}}}
\caption{
Comparison between the X-ray spectrum
of V4743\,Sgr taken on day 742 and a post-SSS nova spectrum
(top) and that of the intermediate polar EX\,Hya (bottom).
The thick grey line in the top panel is a rescaled best-fit model
to an X-ray spectrum of V382\,Vel taken 334 days after discovery
after convolution through the MOS1 response. In the bottom
panel, normalized counts from our MOS1 spectrum and a
\swift/XRT count spectrum are compared.
\label{cmp}}
\end{figure}


\subsection{Discussion of the SSS phase}

This is the first time that such results is extracted
from observations, and
a direct comparison with other works is not possible
at the moment, but it implies that the kinematics
of the ejecta comes from the radiation
supplied by the central source;
{\it suggesting the validity of the equipartition
of energy} \cite{shara2010a}.

We assume an {\it ad hoc}
efficiency of the radiation
to ionization $L_{\rm ion}/L_{\rm rad}=0.41$.
A precise determination of $L_{\rm rad}$ is not simple, but we explore 
plausible values of it.
We make use of the
grid of novae models from \cite{yaron2005a}, to estimate
the mass of the white dwarf, M$_{\rm WD}$,
the temperature of its isothermal core, T$_{\rm WD}$,
and the mass transfer rate, $\dot{M}_{\rm WD}$.
If we take the average of $m_{ej}\approx 4\times 10^{-4}~M_{\sun}$
(assuming $L=10^{38}$ \ergs),
and look for the ``best"-fit with the
$m_{ej}$ in Table 2 of \cite{yaron2005a},
we find that the best- set
of parameters is:
M$_{\rm WD}=0.4$, T$_{\rm WD}=10^{7}$~K and
log$_{10}(\dot{M_{\rm WD}})=-9~(M_{\sun}/yr)$.
That would imply that we are in the presence
of a low-mass ($\lesssim 0.5$ $M_{\sun}$)
WD close binary system
\cite{yaron2005a}.
On the other hand, if we look for parameters with 
$m_{ej}\approx 6\times 10^{-4}~M_{\sun}$,
then
M$_{\rm WD}=0.65$, T$_{\rm WD}=10^{7}$~K and
log$_{10}(\dot{M}_{\rm WD})=-12~(M_{\sun}/yr)$.
However assuming we take our computed luminosity
of $L\approx 6 \times 10^{38}$ \ergs,
then our estimation of $m_{ej}$,
would be $\approx 2\times 10^{-3}~M_{\sun}$
and the extended grid of models \cite{shara2010b}
is required instead.
In that case
M$_{\rm WD}=0.5$, T$_{\rm WD}=3 \times 10^{6}$~K and
$\dot{M}_{\rm WD}=5 \times 10^{-11} ~(M_{\sun}/yr)$,
posing our object at the border
of a very luminous red novae. 
This is certainly an upper limit since
$L\approx 6 \times 10^{38}$ \ergs \sp
is only for epoch 1 and not
for the whole time-interval
of the mass loss.
At the moment
more precise estimations of the luminosity
are required
to characterize the set of physical parameters
of the system \sgr.
More work on that regard might take place
in the near future.

\subsection{Discussion Post SSS phase}
\label{disc:lastobs}

The basic idea of the final stages of nova evolution is
that the ionized ejecta are radiatively cooling. Kinetic
energy is successively converted to radiation via bound-bound
collisional excitations followed by radiative deexcitations.
As discussed in \S\ref{disc:early}, charge exchange may also
be an efficient cooling mechanism. In any way, the typical
spectrum of a radiatively cooling plasma is an emission line
spectrum, and appropriate models are thin-thermal models as,
e.g., implemented in the {\tt apec} module in xspec.

A good example of a spectrum of radiatively cooling ejecta are
post-SSS \chan \space
observations of the nova V382\,Vel, taken on days
268, 334, and 450 after discovery. The first of these spectra was
analyzed by \cite{ness_vel}, and the assumption of a radiatively cooling
plasma proved valid. The other two observations were successfully
modeled with thin-thermal plasma models by \cite{vel_burwitz},
who also found a slow power-law decline from all three observations.

\begin{figure}
\resizebox{10cm}{!}{\includegraphics[angle=0]{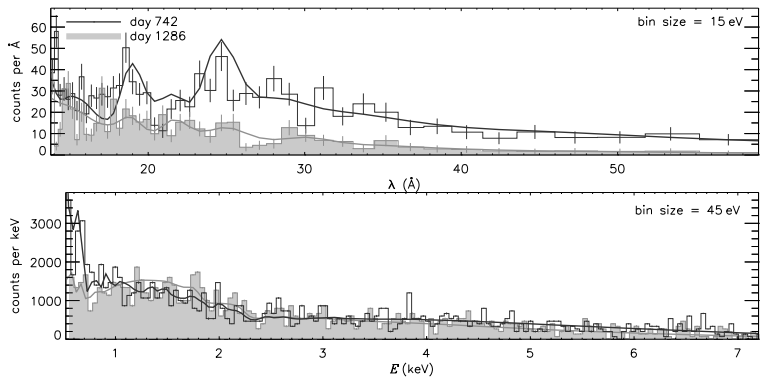}}
\caption{
\xmm/MOS1 spectra taken on days 742 and
1286, plotted with dark and light colors, respectively (see
top left legend). In the top panel, the soft part is shown in
units wavelength,
and in the bottom part the hard part in units keV. Best-fit models
are included with continuous lines.
\label{latecmp}}
\end{figure}

Both post-SSS spectra of V4743\,Sgr are different to V382\,Vel.
In order to compare these spectra, we show in the top panel of Fig.~\ref{cmp}
the \xmm/MOS1 spectrum of V4743\,Sgr taken on day 742 and a rescaled model
that fits the \chan \space
spectrum of V382\,Vel, taken on day 334, after
convolution through the instrumental response of the MOS1 detector.
The spectrum of V382\,Vel contains strong emission lines that are
resolvable with the MOS1 detector, and no significant emission arises
above $\sim 2$\,keV. Meanwhile, our spectrum is dominated by featureless
emission, including a hard tail extending up to 10\,keV.
We were unable to find a satisfactory fit
to the spectrum using an optically thin, thermal plasma model.
Furthermore, no significant fading
of the hard emission can be identified between days 742 and
1286 (see Fig.\ref{latecmp}), while considerable fading was seen
in V382\,Vel (\cite{vel_burwitz}). All evidence we have thus indicates
that the hard component is not typical post-nova emission.

It is also of interest to note that no X-ray emission was seen
in a deep ROSAT observation in 1990, twelve years before outburst.
The nova has thus not returned to quiescence yet. Since the ROSAT band
only covers the energy range 0.1-2.4\,keV, this non-detection only
indicates the absence of soft emission, while the hard tail
could have been present at that time. We thus predict that at least
the soft component will eventually disappear.

\section{Summary and Conclusions}
\label{concl}

We studied the hardness and brightness evolution using eight X-ray
observations of V4743\,Sgr and found three main stages of evolution.

(1) One X-ray spectrum was taken during the first phase and is an
optically thin emission line spectrum which can be modeled by a
thin-thermal plasma model, indicating that collisional processes are
involved. Shock interactions of different components within the ejecta
are the most commonly accepted model, but the presence of strong
magnetic fields could lead to coronal emission that is heated in
magnetically confined loops, similar to stellar coronae. V4743\,Sgr
has been proposed to be an intermediate polar, and we found
supporting evidence (see below), thus magnetic fields may be
present and relevant.

(2) The five high-resolution spectra obtained during the brightest
SSS phase provide the largest amount of detail. The spectrum resembles
a stellar atmosphere, but the absorption lines are blue shifted,
and their complex profiles indicate that we are dealing with
an expanding atmosphere, and the ejecta are inhomogeneous.
Our technique of line
identification,
allow us to identify
lines coming from
H- and He-like ions of C,N,O blueshifted
by $\sim 2500$ \kms, lead us to estimate a mass loss rate of
$\dot{M} \approx (3-5) \times 10^{-4} ~ (\frac{L}{L_{38}}) ~ M_{\sun}/yr$.
The total ejected mass is
$m_{\rm ej} \approx (2-3) \times 10^{-4} ~ M_{\sun}$,
which locate V4743 Sgr., as a low-mass WD CO system.
However due to uncertainties in the intrinsic luminosity
of the source others (extreme?)
positions in the parameter space of the grid of classical
novae models, cannot be ruled out.

(3) The final phase is different from that observed in other novae, for
which radiative cooling of the ionized ejecta has been observed.
While significant fading of emission below $\sim2$\,keV 
is seen, the harder part of the spectrum does not change and
resembles that of the intermediate polar EX\,Hya. ROSAT has
detected no soft X-ray emission before the outburst that
could have swept away material that blocks all soft emission.
As this material replenishes, all emission below 2\,keV might
eventually disappear. This can be tested with further observations.
Meanwhile, the hard emission could stay, which would support
the intermediate polar nature of V4743\,Sgr.

\textsf{acknowledgments}
The author is indebted
to J.-U. Ness and M. Bautista for important discussions, 
carried out in Berlin (AIP, Germany) and Kalamazoo (WMU, EEUU),
very
useful for the realization of this work.
This contribution is partially supported by {\sc ivic} (Venezuela)
project 2013000259.
Also, it was partially supported by ABACUS, CONACyT (Mexico)
grant EDOMEX-2011-C01-165873.



\begin{thebibliography}{35}
\providecommand{\natexlab}[1]{#1}
\providecommand{\url}[1]{\texttt{#1}}
\expandafter\ifx\csname urlstyle\endcsname\relax
  \providecommand{\doi}[1]{doi: #1}\else
  \providecommand{\doi}{doi: \begingroup \urlstyle{rm}\Url}\fi

\bibitem[{Starrfield} and {Swift-Nova-CV group}(2008)]{st08}
S.~{Starrfield} and {Swift-Nova-CV group}.
\newblock {Classical Novae in the Swift Era}.
\newblock In \emph{AAS/High Energy Astrophysics Division}, volume~10 of
  \emph{AAS/High Energy Astrophysics Division}, pages 19.04--+, March 2008.

\bibitem[{Payne-Gaposchkin}(1964)]{novae64}
C.~{Payne-Gaposchkin}.
\newblock \emph{{The galactic novae}}.
\newblock 1964.

\bibitem[{Starrfield} et~al.(2004){Starrfield}, {Timmes}, {Hix}, {Sion},
  {Sparks}, and {Dwyer}]{starr04}
S.~{Starrfield}, F.~X. {Timmes}, W.~R. {Hix}, E.~M. {Sion}, W.~M. {Sparks}, and
  S.~J. {Dwyer}.
\newblock {Surface Hydrogen-burning Modeling of Supersoft X-Ray Binaries: Are
  They Type Ia Supernova Progenitors?}
\newblock \emph{\apjl}, 612:\penalty0 L53--L56, September 2004.
\newblock \doi{10.1086/424513}.

\bibitem[{Krautter} et~al.(1996){Krautter}, {Oegelman}, {Starrfield},
  {Wichmann}, and {Pfeffermann}]{krautter1996a}
J.~{Krautter}, H.~{Oegelman}, S.~{Starrfield}, R.~{Wichmann}, and
  E.~{Pfeffermann}.
\newblock {ROSAT X-Ray Observations of Nova V1974 Cygni: The Rise and Fall of
  the Brightest Supersoft X-Ray Source}.
\newblock \emph{\apj}, 456:\penalty0 788--+, January 1996.
\newblock \doi{10.1086/176697}.

\bibitem[{Balman} et~al.(1998{\natexlab{a}}){Balman}, {Krautter}, and
  {Oegelman}]{balman1998a}
S.~{Balman}, J.~{Krautter}, and H.~{Oegelman}.
\newblock {The X-Ray Spectral Evolution of Classical Nova V1974 Cygni 1992: A
  Reanalysis of the ROSAT Data}.
\newblock \emph{\apj}, 499:\penalty0 395--+, May 1998{\natexlab{a}}.
\newblock \doi{10.1086/305600}.

\bibitem[{Lloyd} et~al.(1992){Lloyd}, {O'Brien}, {Bode}, {Predehl}, {Schmitt},
  {Truemper}, {Watson}, and {Pounds}]{lloyd92}
H.~M. {Lloyd}, T.~J. {O'Brien}, M.~F. {Bode}, P.~{Predehl}, J.~H.~M.~M.
  {Schmitt}, J.~{Truemper}, M.~G. {Watson}, and K.~A. {Pounds}.
\newblock {X-ray detection of Nova Herculis 1991 five days after optical
  outburst}.
\newblock \emph{\nat}, 356:\penalty0 222--224, March 1992.
\newblock \doi{10.1038/356222a0}.

\bibitem[{Orio} et~al.(2001){Orio}, {Parmar}, {Benjamin}, {Amati}, {Frontera},
  {Greiner}, {{\"O}gelman}, {Mineo}, {Starrfield}, and {Trussoni}]{Orio2001}
M.~{Orio}, A.~{Parmar}, R.~{Benjamin}, L.~{Amati}, F.~{Frontera}, J.~{Greiner},
  H.~{{\"O}gelman}, T.~{Mineo}, S.~{Starrfield}, and E.~{Trussoni}.
\newblock {The X-ray emission from Nova V382 Velorum - I. The hard component
  observed with BeppoSAX}.
\newblock \emph{\mnras}, 326:\penalty0 L13--L18, September 2001.
\newblock \doi{10.1046/j.1365-8711.2001.04448.x}.

\bibitem[{Ness} et~al.(2007){Ness}, {Schwarz}, {Retter}, {Starrfield},
  {Schmitt}, {Gehrels}, {Burrows}, and {Osborne}]{swnovae}
{J.-U.} {Ness}, G.~J. {Schwarz}, A.~{Retter}, S.~{Starrfield}, J.~H.~M.~M.
  {Schmitt}, N.~{Gehrels}, D.~{Burrows}, and J.~P. {Osborne}.
\newblock {Swift X-Ray Observations of Classical Novae}.
\newblock \emph{\apj}, 663:\penalty0 505--515, July 2007.
\newblock \doi{10.1086/518084}.

\bibitem[{Page} et~al.(2010){Page}, {Osborne}, {Evans}, {Wynn}, {Beardmore},
  {Starling}, {Bode}, {Ibarra}, {Kuulkers}, {Ness}, and {Schwarz}]{page09}
K.~L. {Page}, J.~P. {Osborne}, P.~A. {Evans}, G.~A. {Wynn}, A.~P. {Beardmore},
  R.~L.~C. {Starling}, M.~F. {Bode}, A.~{Ibarra}, E.~{Kuulkers}, {J.-U.}
  {Ness}, and G.~J. {Schwarz}.
\newblock {Swift observations of the X-ray and UV evolution of V2491 Cyg (Nova
  Cyg 2008 No. 2)}.
\newblock \emph{\mnras}, 401:\penalty0 121--130, January 2010.
\newblock \doi{10.1111/j.1365-2966.2009.15681.x}.

\bibitem[{Tsujimoto} et~al.(2009){Tsujimoto}, {Takei}, {Drake}, {Ness}, and
  {Kitamoto}]{v458}
M.~{Tsujimoto}, D.~{Takei}, J.~J. {Drake}, {J.-U.} {Ness}, and S.~{Kitamoto}.
\newblock {X-Ray Spectroscopy of the Classical Nova V458 Vulpeculae with
  Suzaku}.
\newblock \emph{\pasj}, 61:\penalty0 69--+, January 2009.

\bibitem[{Ness} et~al.(2008){Ness}, {Schwarz}, {Starrfield}, {Osborne}, {Page},
  {Beardmore}, {Wagner}, and {Woodward}]{ness_v723}
{J.-U.} {Ness}, G.~{Schwarz}, S.~{Starrfield}, J.~P. {Osborne}, K.~L. {Page},
  A.~P. {Beardmore}, R.~M. {Wagner}, and C.~E. {Woodward}.
\newblock {V723 CASSIOPEIA Still on in X-Rays a Bright Super Soft Source 12
  Years after Outburst}.
\newblock \emph{\aj}, 135:\penalty0 1328--1333, April 2008.
\newblock \doi{10.1088/0004-6256/135/4/1328}.

\bibitem[{Sala} and {Hernanz}(2005)]{sala05a}
G.~{Sala} and M.~{Hernanz}.
\newblock {Envelope models for the supersoft X-ray emission of V1974 Cyg}.
\newblock \emph{\aap}, 439:\penalty0 1057--1060, September 2005.
\newblock \doi{10.1051/0004-6361:20042587}.

\bibitem[{Ness} et~al.(2009){Ness}, {Drake}, {Beardmore}, {Boyd}, {Bode},
  {Brady}, {Evans}, {Gaensicke}, {Kitamoto}, {Knigge}, {Miller}, {Osborne},
  {Page}, {Rodriguez-Gil}, {Schwarz}, {Staels}, {Steeghs}, {Takei},
  {Tsujimoto}, {Wesson}, and {Zijlstra}]{ness09}
{J.-U.} {Ness}, J.~J. {Drake}, A.~P. {Beardmore}, D.~{Boyd}, M.~F. {Bode},
  S.~{Brady}, P.~A. {Evans}, B.~T. {Gaensicke}, S.~{Kitamoto}, C.~{Knigge},
  I.~{Miller}, J.~P. {Osborne}, K.~L. {Page}, P.~{Rodriguez-Gil}, G.~{Schwarz},
  B.~{Staels}, D.~{Steeghs}, D.~{Takei}, M.~{Tsujimoto}, R.~{Wesson}, and
  A.~{Zijlstra}.
\newblock {Swift X-Ray and Ultraviolet Monitoring of the Classical Nova V458
  Vul (Nova Vul 2007)}.
\newblock \emph{\aj}, 137:\penalty0 4160--4168, May 2009.
\newblock \doi{10.1088/0004-6256/137/5/4160}.

\bibitem[{van Rossum} and {Ness}(2010)]{vanrossumness09}
D.~R. {van Rossum} and {J.-U.} {Ness}.
\newblock {Expanding atmosphere models for SSS spectra of novae}.
\newblock \emph{Astronomische Nachrichten}, 331:\penalty0 175--+, 2010.
\newblock \doi{10.1002/asna.200911321}.

\bibitem[{Ness} et~al.(2003){Ness}, {Starrfield}, {Burwitz}, {Wichmann},
  {Hauschildt}, {Drake}, {Wagner}, {Bond}, {Krautter}, {Orio}, {Hernanz},
  {Gehrz}, {Woodward}, {Butt}, {Mukai}, {Balman}, and {Truran}]{ness2003a}
{J.-U.} {Ness}, S.~{Starrfield}, V.~{Burwitz}, R.~{Wichmann}, P.~{Hauschildt},
  J.~J. {Drake}, R.~M. {Wagner}, H.~E. {Bond}, J.~{Krautter}, M.~{Orio},
  M.~{Hernanz}, R.~D. {Gehrz}, C.~E. {Woodward}, Y.~{Butt}, K.~{Mukai},
  S.~{Balman}, and J.~W. {Truran}.
\newblock {A Chandra Low Energy Transmission Grating Spectrometer Observation
  of V4743 Sagittarii: A Supersoft X-Ray Source and a Violently Variable Light
  Curve}.
\newblock \emph{\apjl}, 594:\penalty0 L127--L130, September 2003.
\newblock \doi{10.1086/378664}.

\bibitem[{Scargle}(2004)]{scargle2004a}
J.~D. {Scargle}.
\newblock {Data Analysis Through Segmentation: Bayesian Blocks and Beyond}.
\newblock In \emph{Bulletin of the American Astronomical Society}, volume~36 of
  \emph{Bulletin of the American Astronomical Society}, pages 1200--+, August
  2004.

\bibitem[{Grevesse} et~al.(1996){Grevesse}, {Noels}, and
  {Sauval}]{grevesse1996a}
N.~{Grevesse}, A.~{Noels}, and A.~J. {Sauval}.
\newblock {Standard Abundances}.
\newblock In \emph{Astronomical Society of the Pacific Conference Series}, page
  117, 1996.

\bibitem[{Kallman} and {Bautista}(2001)]{kallman2001a}
T.~{Kallman} and M.~{Bautista}.
\newblock {Photoionization and High-Density Gas}.
\newblock \emph{\apjs}, 133:\penalty0 221--253, March 2001.

\bibitem[{O'Brien} et~al.(1994){O'Brien}, {Lloyd}, and {Bode}]{obrien94}
T.~J. {O'Brien}, H.~M. {Lloyd}, and M.~F. {Bode}.
\newblock {An interacting winds model for the X-ray emission from V838 Her
  (Nova Herculis 1991).}
\newblock \emph{\mnras}, 271:\penalty0 155--160, November 1994.

\bibitem[{Smith} et~al.(2001){Smith}, {Brickhouse}, {Liedahl}, and
  {Raymond}]{smith01}
R.~K. {Smith}, N.~S. {Brickhouse}, D.~A. {Liedahl}, and J.~C. {Raymond}.
\newblock {Collisional Plasma Models with APEC/APED: Emission-Line Diagnostics
  of Hydrogen-like and Helium-like Ions}.
\newblock \emph{\apjl}, 556:\penalty0 L91--L95, August 2001.
\newblock \doi{10.1086/322992}.

\bibitem[{Balman} et~al.(1998{\natexlab{b}}){Balman}, {Krautter}, and
  {Oegelman}]{balm98}
S.~{Balman}, J.~{Krautter}, and H.~{Oegelman}.
\newblock {The X-Ray Spectral Evolution of Classical Nova V1974 Cygni 1992: A
  Reanalysis of the ROSAT Data}.
\newblock \emph{\apj}, 499:\penalty0 395--+, May 1998{\natexlab{b}}.
\newblock \doi{10.1086/305600}.

\bibitem[{Lyke} et~al.(2002){Lyke}, {Kelley}, {Gehrz}, and {Woodward}]{lyke}
J.~E. {Lyke}, M.~S. {Kelley}, R.~D. {Gehrz}, and C.~E. {Woodward}.
\newblock {Free-Free Turnover in Nova V4743 Sgr 2002 \#3}.
\newblock In \emph{Bulletin of the American Astronomical Society}, volume~34 of
  \emph{Bulletin of the American Astronomical Society}, pages 1161--+, December
  2002.

\bibitem[{Kang} et~al.(2006){Kang}, {Retter}, {Liu}, and {Richards}]{kang06}
T.~W. {Kang}, A.~{Retter}, A.~{Liu}, and M.~{Richards}.
\newblock {Nova V4743 Sagittarii 2002: An Intermediate Polar Candidate}.
\newblock \emph{\aj}, 132:\penalty0 608--613, August 2006.
\newblock \doi{10.1086/505174}.

\bibitem[{Dobrotka} and {Ness}(2010)]{dobrness09}
A.~{Dobrotka} and {J.-U.} {Ness}.
\newblock {Multifrequency nature of the 0.75 mHz feature in the X-ray light
  curves of the nova V4743 Sgr}.
\newblock \emph{\mnras}, 405:\penalty0 2668--2682, July 2010.
\newblock \doi{10.1111/j.1365-2966.2010.16654.x}.

\bibitem[{Ness} and {Starrfield}(2009)]{rsophshock}
{J.-U.} {Ness} and S.~{Starrfield}.
\newblock {The X-ray Spectra of the Shock Systems in RS Oph}.
\newblock In {S.~J.~Murphy \& M.~S.~Bessell}, editor, \emph{Astronomical
  Society of the Pacific Conference Series}, volume 404 of \emph{Astronomical
  Society of the Pacific Conference Series}, pages 77--+, August 2009.

\bibitem[{O'Brien} et~al.(1992){O'Brien}, {Bode}, and {Kahn}]{obrien92}
T.~J. {O'Brien}, M.~F. {Bode}, and F.~D. {Kahn}.
\newblock {Models for the remnants of recurrent novae. III - Comparison with
  the X-ray observations of RS Ophiuchi (1985)}.
\newblock \emph{\mnras}, 255:\penalty0 683--693, April 1992.

\bibitem[{Bode}(2004)]{bode04}
M.~F. {Bode}.
\newblock {The Evolution of Nova Ejecta}.
\newblock In {M.~Meixner, J.~H.~Kastner, B.~Balick, \& N.~Soker}, editor,
  \emph{Asymmetrical Planetary Nebulae III: Winds, Structure and the
  Thunderbird}, volume 313 of \emph{Astronomical Society of the Pacific
  Conference Series}, pages 504--+, July 2004.

\bibitem[{Lloyd} et~al.(1995){Lloyd}, {O'Brien}, and {Bode}]{lloyd95}
H.~M. {Lloyd}, T.~J. {O'Brien}, and M.~F. {Bode}.
\newblock {Interacting Winds in Classical Nova Outbursts}.
\newblock \emph{\apss}, 233:\penalty0 317--321, November 1995.
\newblock \doi{10.1007/BF00627366}.

\bibitem[{Garc{\'{\i}}a} et~al.(2005){Garc{\'{\i}}a}, {Mendoza}, {Bautista},
  {Gorczyca}, {Kallman}, and {Palmeri}]{garcia2005a}
J.~{Garc{\'{\i}}a}, C.~{Mendoza}, M.~A. {Bautista}, T.~W. {Gorczyca}, T.~R.
  {Kallman}, and P.~{Palmeri}.
\newblock {K-Shell Photoabsorption of Oxygen Ions}.
\newblock \emph{\apjs}, 158:\penalty0 68--79, May 2005.
\newblock \doi{10.1086/428712}.

\bibitem[{Hartmann} and {Heise}(1997)]{hartheis97}
H.~W. {Hartmann} and J.~{Heise}.
\newblock {Hot high-gravity NLTE model atmospheres as soft X-ray sources.}
\newblock \emph{\aap}, 322:\penalty0 591--597, June 1997.

\bibitem[{Shara} et~al.(2010{\natexlab{a}}){Shara}, {Yaron}, {Prialnik}, and
  {Kovetz}]{shara2010a}
M.~M. {Shara}, O.~{Yaron}, D.~{Prialnik}, and A.~{Kovetz}.
\newblock {Non-equipartition of Energy, Masses of Nova Ejecta, and Type Ia
  Supernovae}.
\newblock \emph{\apjl}, 712:\penalty0 L143--L147, April 2010{\natexlab{a}}.
\newblock \doi{10.1088/2041-8205/712/2/L143}.

\bibitem[{Yaron} et~al.(2005){Yaron}, {Prialnik}, {Shara}, and
  {Kovetz}]{yaron2005a}
O.~{Yaron}, D.~{Prialnik}, M.~M. {Shara}, and A.~{Kovetz}.
\newblock {An Extended Grid of Nova Models. II. The Parameter Space of Nova
  Outbursts}.
\newblock \emph{\apj}, 623:\penalty0 398--410, April 2005.
\newblock \doi{10.1086/428435}.

\bibitem[{Shara} et~al.(2010{\natexlab{b}}){Shara}, {Yaron}, {Prialnik},
  {Kovetz}, and {Zurek}]{shara2010b}
M.~M. {Shara}, O.~{Yaron}, D.~{Prialnik}, A.~{Kovetz}, and D.~{Zurek}.
\newblock {An Extended Grid of Nova Models. III. Very Luminous, Red Novae}.
\newblock \emph{ArXiv e-prints}, September 2010{\natexlab{b}}.

\bibitem[{Ness} et~al.(2005){Ness}, {Starrfield}, {Jordan}, {Krautter}, and
  {Schmitt}]{ness_vel}
{J.-U.} {Ness}, S.~{Starrfield}, C.~{Jordan}, J.~{Krautter}, and J.~H.~M.~M.
  {Schmitt}.
\newblock {An X-ray emission-line spectrum of Nova V382Velorum 1999}.
\newblock \emph{\mnras}, 364:\penalty0 1015--1024, December 2005.
\newblock \doi{10.1111/j.1365-2966.2005.09664.x}.

\bibitem[{Burwitz} et~al.(2002){Burwitz}, {Starrfield}, {Krautter}, and
  {Ness}]{vel_burwitz}
V.~{Burwitz}, S.~{Starrfield}, J.~{Krautter}, and {J.-U.} {Ness}.
\newblock {Chandra ACIS-I and LETGS X-ray observations of Nova 1999 Velorum
  (V382 Vel)}.
\newblock In {M.~Hernanz \& J.~Jos{\'e}}, editor, \emph{Classical Nova
  Explosions}, volume 637 of \emph{American Institute of Physics Conference
  Series}, pages 377--380, November 2002.
\newblock \doi{10.1063/1.1518233}.

\end{thebibliography}




\end{document}